# Critically coupled zeroth-order resonance for ultrathin nonlinear photonics

Hae-Seok Jeong[1,†], Soon-Jae Lee[1,†], Young-Ho Jin[1], Su-Hyun Gong[1,*] and Q-Han Park[1,*]

[1]Department of Physics, Korea University, Seoul, 02841, Republic of Korea

**Abstract**

Ultrathin active materials are essential for compact nonlinear and quantum photonic devices, yet no general principle exists to link their optical constants to the cavity designs required for simultaneous field buildup and reflection suppression. Consequently, achieving extreme optical confinement currently relies on trial-and-error optimization for every new material. Here, we establish a design rule for metal-backed cavities that maximizes light-matter interaction by ensuring the simultaneous satisfaction of zeroth-order resonance and critical coupling. We derive a closed-form analytical condition that partitions the (n, k) plane into critically coupled, over-coupled, and under-coupled regimes, each mapping to a specific minimal architecture. The critical curve admits a three-layer open cavity, the over-coupled region a closed cavity with a semi-transparent top mirror, and the under-coupled region a spacer-assisted geometry. For low-loss materials, the closed cavity spatially separates dissipation from field accumulation, allowing the quality factor to be controlled by the external mirror rather than intrinsic medium absorption. We validate this framework with 3R-$MoS_2$, demonstrating a second-harmonic enhancement of $1.19 \times 10^5$ relative to a monolayer, accompanied by the near-complete suppression of reflected pump waves. These results provide a universal framework for efficient light-matter interaction in ultrathin nonlinear and quantum photonics.

## Introduction

Nanophotonic cavities with active layers only a few to tens of nanometers thick provide a route to compact nonlinear and quantum photonic devices[1]. However, concentrating optical fields in such deep-subwavelength regions remains a central challenge, especially as progress in atomically thin and few-layer materials such as transition metal dichalcogenides[2–4] and layered perovskites[5,6] offers exceptionally large nonlinear susceptibilities and quantum-light functionality compatible with planar integration[7,8].

The fundamental difficulty in these ultrathin systems is that the conditions for strong field buildup and reflection suppression are not naturally aligned. In a metal-backed, one-port geometry where transmission is eliminated, suppressing reflection requires a precise balance between external coupling and internal loss. However, this internal loss typically limits the cavity quality factor and the resulting field intensity inside the active layer. Consequently, the simplest cavity geometries fail to simultaneously achieve maximum field accumulation and the vanishing reflection required for high-efficiency operation.

Existing platforms, including distributed Bragg reflectors[9–11], Fabry–Pérot cavities[12], plasmonic nanoantennas[13], photonic crystal slabs[14–17] and microcavities[18], and nonlinear metasurfaces[19–22], generate large field enhancements, but each is restricted by structural features that must be bespoke-optimized for every specific material and wavelength. These platform designs often rely on complex lateral patterning or couple fields through highly localized hot spots, lacking a universal design rule. While critical coupling, the condition where the external coupling rate $\gamma_e$ equals the internal loss rate $\gamma_i$, serves as a guide for reflection suppression[23,24], applying it to the ultrathin regime requires precise destructive interference achieved by managing the linear round-trip phase through zeroth-order resonance. In this regime, the minimal propagation phase accumulated within the film is offset by the negative reflection phase at the metal interface, satisfying the linear round-trip interference condition for the pump field. This resonance mechanism is fundamentally distinct from the nonlinear phase matching typically required between fundamental and harmonic waves. Because the active layer, only tens of nanometers thick, is orders of magnitude thinner than the second-harmonic coherence length, conventional phase-matching constraints are bypassed, allowing the conversion efficiency to be governed entirely by the cavity-enhanced local field.

Furthermore, while ultrathin perfect absorbers have realized critical coupling in vertical geometries[25–27], they typically rely on a single lossy material to provide both field confinement and the dissipation required for vanishing reflection[25,26]. This configuration, while effective for maximizing absorption, is suboptimal for nonlinear processes like second-harmonic generation (SHG), which are driven by local field intensity rather than absorbed power. To maximize such processes, a spatial separation of dissipation from field accumulation is required[28].

Here, we establish a general cavity design framework that resolves this limitation. By simultaneously requiring zeroth-order resonance and critical coupling, we derive a closed-form analytical criterion that partitions the ($n$, $k$) plane into critically coupled, over-coupled, and under-coupled regimes, each mapping to a specific minimal architecture. The critical curve admits a three-layer open cavity, the over-coupled region a closed cavity with a semi-transparent top mirror, and the under-coupled region a spacer-assisted geometry. We implement this framework experimentally using 3R-$MoS_2$, whose high refractive index, low optical loss and thickness-independent broken inversion symmetry make it well suited for probing the over-coupled nonlinear regime. Using lithography-free vertical stacks, we compare an open cavity that isolates zeroth-order field confinement with a closed cavity that brings the same active layer to critical coupling. This direct comparison reveals how impedance matching reshapes the nonlinear response while spatially separating field accumulation from dissipation. Because the design rule depends only on the optical constants of the active layer and the metal, the same framework prescribes architectures for other ultrathin materials across visible to telecommunication wavelengths.

## Results

### Critical coupling in zeroth-order cavities

We first establish the conditions for critical coupling within an ultrathin film supported by a metallic substrate, a minimal architecture shown in Fig. 1a. This is a one-port system in which the metal substrate eliminates transmission and the air interface is the only external coupling channel. Critical coupling is achieved when the reflectance vanishes and the incident wave is fully coupled into the cavity rather than reflected. Setting the reflectance to zero yields two independent conditions on the film thickness *d*, corresponding to amplitude matching and phase matching for zeroth-order resonance (Supplementary Note 1):

$$d_{amp} = -\frac{\ln|z|}{2k_0 k} \qquad (1a)$$
$$d_{phase} = \frac{\arg(z)}{2k_0 n} \qquad (1b)$$

where $z \equiv -r_{01}/r_{12}$ is determined by the Fresnel coefficients at the air-film and film-metal interfaces, which depend on the film's complex refractive index $n + ik$ and the metal's $\tilde{n}_M$, and $k_0 = 2\pi/\lambda$. Equation (1b) is the zeroth-order phase condition ($\Delta\phi = 0$). It fixes the thickness at which the propagation phase accumulated within the film is offset by the relative reflection phase of the air-film and film-metal interfaces, dominated by the negative reflection phase at the metal interface, selecting the ultrathin branch of the resonance condition. Equation (1a) is the independent amplitude condition. It requires round-trip absorption to compensate for the reflectivity imbalance between the two interfaces, so that the effective reflected amplitudes have equal magnitude. The reflected field vanishes only when this amplitude balance is satisfied at a thickness that also satisfies the resonance phase condition. Critical coupling in this geometry therefore requires $d_{\mathrm{amp}} = d_{\mathrm{phase}}$, corresponding to the simultaneous satisfaction of two independent thickness requirements derived from the single complex variable $z$. These are a phase-condition thickness that establishes zeroth-order resonance and an amplitude-condition thickness that balances external coupling with internal loss. The two thicknesses are not in general equal, and their mismatch determines whether the cavity is over-coupled, critically coupled, or under-coupled. In a conventional Fabry–Pérot cavity, the round-trip phase condition[29,30] $2k_0 n(\lambda)d + \phi_{bot} + \phi_{top} = 2m\pi$ requires $m \geq 1$, hence $d \geq \lambda/2n$. The negative reflection phase $\phi_{bot} < 0$ at the metal interface offsets part of the propagation phase, allowing the condition to be satisfied at $m = 0$ with $d$ far below the half-wavelength limit.[25–27,31] This sub-half-wavelength operation is the structural basis for the ultrathin design framework presented here. Equating both expressions for $d$ yields the master equation for critical coupling.

$$-n \ln|z| = k \arg(z) \qquad (2)$$

Because $z$ depends only on the refractive indices of the film $(n, k)$ and the complex refractive index of the metal substrate ($\tilde{n}_M$), not on the thickness, this equation constrains the permissible material parameters to a unique curve in the $(n, k)$ plane for a given metal and wavelength (Fig. 1c). Materials whose optical constants lie on this curve can reach critical coupling in the simplest three-layer geometry. All other materials require structural modification.

The physical content of Eq. (2) expresses the condition for critical coupling of the zeroth-order resonance. The phase condition first determines the resonant thickness, $d = d_{\mathrm{phase}}$, and the remaining question is whether the amplitude balance is also satisfied at this same thickness. Within the temporal coupled-mode theory of this one-port cavity[23,24], the external coupling rate $\gamma_e$ (through the air interface) and the internal loss rate $\gamma_i$ (material and metal absorption) determine the on-resonance reflectance $R = \left|\frac{\gamma_e - \gamma_i}{\gamma_e + \gamma_i}\right|^2$ (Supplementary Note 1). The reflectance vanishes when these two rates are balanced, which represents the defining condition of critical coupling. Both this rate-based criterion and the amplitude condition (Eq. 1a) describe the same physical mechanism for the vanishing of the reflected field. In the three-layer system, the resonant phase condition fixes $d$, which in turn fixes the ratio $\gamma_e/\gamma_i$. The critical coupling curve therefore represents the set of optical constants for which this balance is naturally satisfied without any additional structural degree of freedom. Materials lying below the curve are over-coupled ($\gamma_e > \gamma_i$), while those above are under-coupled ($\gamma_e < \gamma_i$); both regimes require structural modification to reach the balance.

Each off-curve regime is matched to a specific minimal four-layer architecture (Fig. 1b) that restores critical coupling without re-engineering the active medium itself. For over-coupled materials, adding a semi-transparent top metal layer reduces $\gamma_e$ by attenuating the external coupling. Although a metallic layer is usually associated with additional absorption, here it also acts as an anti-reflection element for the fundamental wave through destructive interference of the reflected field. The required thickness follows from the critical coupling condition as $d_t \approx \frac{1}{4k_0 k_M}\ln(\frac{\gamma_e}{\gamma_i})$ (Supplementary Note 1). For under-coupled materials, a lossless dielectric spacer (e.g., $Al_2O_3$) inserted between the film and the metal substrate decouples the phase and amplitude conditions, allowing independent tuning of the resonance thickness without altering the absorptive properties of the film.

The master equation Eq. (2) further distinguishes two physically distinct operating regimes. Large-extinction materials lie on or near the critical curve, where the amplitude condition is satisfied within a single round-trip and the cavity behaves as a low-Q single-pass absorber. Low-extinction nonlinear materials, by contrast, lie deep in the over-coupled regime, where the three-layer geometry cannot satisfy the amplitude condition alone; adding a semi-transparent top mirror reduces $\gamma_e$ until it matches $\gamma_i$, producing a relatively high-Q mode in which the field accumulates over more round-trips. For nonlinear photonics, the relevant figure of merit is the field intensity experienced by the nonlinear polarization rather than the absorbed power, which selects the high-Q (over-coupled) regime. This regime contains most low-loss nonlinear materials of interest, and the closed-cavity geometry analyzed in the remainder of this work brings such materials to critical coupling.

The practical use of this framework is illustrated in Fig. 1c. For a given material at its operating wavelength, one locates the point $(n, k)$ in the map and identifies the applicable regime. A three-layer open cavity applies to points on the curve, a closed cavity with a top Au mirror to over-coupled points below the curve, and a spacer-assisted geometry to under-coupled points above it. For example, 3R-$MoS_2$ at 910 nm falls far below the 910 nm curve, placing it deep in the over-coupled regime and prescribing the closed cavity architecture implemented in this work. This prescriptive use of the map eliminates trial-and-error optimization and provides a clear design guideline for any candidate nonlinear material.

Representative examples spanning different regions of the $(n, k)$ map are provided in Supplementary Figs. S4–S7.[32–36]

## Zeroth-order cavity implementation with 3R-$MoS_2$

We now apply this framework to 3R-$MoS_2$, whose combination of broken inversion symmetry and high refractive index makes it a representative platform for ultrathin nonlinear photonics. Unlike the 2H polytype, in which inversion symmetry is restored in even-numbered layers and suppresses bulk second-order nonlinearity, the 3R polytype preserves broken inversion symmetry irrespective of thickness[37–39], allowing the SHG response to scale constructively with volume. Near the operating wavelength and below the excitonic resonance at 910 nm, 3R-$MoS_2$ exhibits a high refractive index (n = 4.4) together with low absorption[3,40,41]. In the $(n, k)$ map of Fig. 1c, this combination places 3R-$MoS_2$ deep within the over-coupled regime, with a three-layer minimum reflectance exceeding 0.84 at 918 nm. The framework therefore prescribes the closed cavity architecture with a semi-transparent top Au mirror as the minimal structure required to bring 3R-$MoS_2$ to critical coupling.

The closed cavity realizes this critical coupling through a spatial separation of field confinement and dissipation. The field is concentrated within the low-absorption 3R-$MoS_2$ film, while the dissipation required for critical coupling is supplied by the top Au layer. This spatial separation allows the intracavity field to build up over many round trips, producing a quality factor substantially higher than in any geometry where a single material provides both functions. We implement this closed cavity alongside a three-layer open cavity that satisfies only the phase condition of Eq. (1b). This pairing isolates the contribution of zeroth-order resonance alone from that of critical coupling.

Figs. 2a and 2b illustrate the two cavity configurations. In the open cavity (Air/3R-$MoS_2$/$Al_2O_3$/Au), the fundamental wave (FW) acquires a negative phase shift upon reflection from the Au substrate, which compensates for the propagation phase and establishes the zeroth-order resonance at a thickness far below $\lambda/2n$. The closed cavity (Air/Au/$Al_2O_3$/3R-$MoS_2$/$Al_2O_3$/Au) adds a semi-transparent top Au layer that introduces a second negative phase contribution and provides the impedance matching required for critical coupling. In both configurations, thin $Al_2O_3$ spacers (7 nm) deposited by atomic layer deposition protect the metal interfaces from thermal damage during optical pumping and enable fine-tuning of the interference phase (Supplementary Fig. S8).

The calculated wavelength–thickness dependence of the total round-trip phase is summarized in Figs. 2c and 2d for the open and closed cavities, respectively. Across the fundamental wavelength range of 800–1050 nm, the zeroth-order resonance condition is satisfied for 3R-$MoS_2$ thicknesses of only tens of nanometers, well below the half-wavelength Fabry–Pérot limit marked by the white dashed line in Figs 2e and 2f. Because the negative reflection phase at the metal interface remains nearly constant over this spectral range, the resonance wavelength scales linearly with thickness in both configurations. The corresponding numerically calculated SHG intensity maps (Figs. 2e and 2f) confirm that regions of enhanced SHG coincide precisely with the zeroth-order phase condition.

A distinctive feature of the zeroth-order mode is its nodeless field profile, which maximizes the spatial overlap between the confined fundamental and the nonlinear medium. Because the cavity is resonant only at the fundamental wavelength, the generated second harmonic propagates freely without back-conversion losses that can limit doubly resonant architectures. The fabricated open and closed cavity structures are confirmed by cross-sectional transmission electron microscopy (TEM), which clearly resolves the vertically stacked layer sequences (Supplementary Fig. S9). The 3R-$MoS_2$ films were mechanically exfoliated and transferred onto the substrates, with thicknesses controlled and verified by atomic force microscopy (Supplementary Fig. S10).

**Broadband nonlinear response of the open cavity**

The SHG intensity maps in Fig. 2e predict that zeroth-order resonance alone, even without critical coupling, can produce strong SHG enhancement within the 3R-$MoS_2$ layer. We verify this experimentally using the open cavity as a reference structure that isolates the effect of zeroth-order resonance alone, where only the round-trip phase condition (Eq. 1b) is satisfied while the amplitude condition (Eq. 1a) remains unfulfilled because $\gamma_e$ far exceeds $\gamma_i$ in the 3R-$MoS_2$/$Al_2O_3$/Au geometry.

Fig. 3a compares the SHG spectra of an open cavity incorporating a 27 nm-thick 3R-$MoS_2$ layer with two references, the same-thickness film on $SiO_2$ and a monolayer $MoS_2$ on $SiO_2$, measured under identical conditions. At 910 nm, the open cavity yields an SHG intensity 6.81 $\times 10^4$ times that of the monolayer, compared with only 13-fold for the same film on $SiO_2$. This four-orders-of-magnitude contrast confirms that the enhancement arises from cavity-enabled field confinement rather than from volume scaling, establishing zeroth-order resonance alone as a significant enhancement mechanism even without critical coupling.

The excitation wavelength dependence (Fig. 3b) reveals that the enhancement persists across nearly the entire pump range, with SHG intensities approximately three orders of magnitude above the bare film reference. The broad linewidth (FWHM = 44.0 ± 6.12 nm) is consistent with the low-quality factor expected for the over-coupled regime, where $\gamma_e$ dominates over $\gamma_i$. Experimentally measured spectra show excellent agreement with numerical simulations, confirming that the nonlinear response tracks the zeroth-order phase condition predicted in Fig. 2e.

As the film thickness increases (Fig. 3c), the zeroth-order resonance shifts toward longer wavelengths, but the SHG maximum is modulated by the susceptibility spectrum of 3R-$MoS_2$, which peaks near 910–920 nm (Supplementary Fig. S11). Because the open cavity resonance is broad, it does not impose a narrow spectral filter but instead provides a wide window of field confinement. For thinner samples, the SHG maximum is therefore pinned to the susceptibility peak rather than tracking the cavity resonance. FWHM values of 44–123 nm across multiple thicknesses underscore the high tolerance of the open cavity to thickness variations (Supplementary Fig. S12). The enhancement achieved in the open cavity originates entirely from field confinement without impedance matching, leaving critical coupling to be realized in the closed geometry.

**Critical coupling enhanced nonlinear response in the closed cavity**

The closed cavity (Au/$Al_2O_3$/3R-$MoS_2$/$Al_2O_3$/Au) satisfies both conditions simultaneously. The top Au layer reduces $\gamma_e$ to match $\gamma_i$, thereby suppressing reflection at the critical-coupling condition and fulfilling Eq. (1a) at the same thickness where Eq. (1b) establishes the zeroth-order resonance. Because the field now circulates within the low-absorption 3R-$MoS_2$ while the dissipation occurs at the top Au, the quality factor is substantially higher than in the open cavity.

Fig. 4a maps the reflectance as a function of wavelength and top Au thickness. At 910 nm, the reflectance vanishes for a top Au thickness of 28 nm, in precise agreement with the value predicted by the critical coupling framework (Supplementary Note 1). As shown analytically in Supplementary Note 1, critical coupling simultaneously achieves zero reflectance and maximum intracavity stored energy, so the same 28 nm thickness also maximizes the driving field for nonlinear conversion. When the top Au is fixed at 28 nm and the 3R-$MoS_2$ thickness is varied, the calculated reflectance map (Fig. 4b) shows the dip red-shifting linearly while remaining deeply suppressed. Experimental reflectance measurements (Fig. 4c) closely reproduce these trends, confirming that the resonance is tuned by a single geometric parameter without lateral patterning. The deep suppression persists across the entire thickness range, demonstrating that once the top Au impedance is optimized, the critical coupling condition is robust against variations in resonance wavelength.

The impact on the nonlinear response is demonstrated in Fig. 4d for a representative closed cavity sample with a 3R-$MoS_2$ thickness of 54.8 nm. The SHG maximum coincides precisely with the reflectance minimum at 910 nm, providing direct evidence that critical coupling maximizes the nonlinear response. This synchronization is reproduced across all measured samples (Supplementary Fig. S13), with the SHG peak consistently tracking the reflectance dip as the resonance shifts from 820 to 1030 nm over the thickness range of 42.9 to 71.7 nm.

The SHG enhancement factors, normalized to the monolayer response, are summarized in Fig. 4e. The closed cavity consistently yields factors of $10^4$ to $10^5$, with a peak value of $1.19 \times 10^5$ at a 3R-$MoS_2$ thickness of 54.8 nm, where the cavity resonance overlaps the intrinsic susceptibility peak near 910 nm. In comparison, unstructured films on $SiO_2$ of the same thickness range show enhancements limited to $10$–$10^2$. The SHG peak wavelength scales linearly with the 3R-$MoS_2$ thickness (Fig. 4f), with narrower spectral linewidths (FWHM = 11–27 nm) than the open cavity counterparts, consistent with the higher quality factor achieved through impedance matching. The nearly twofold improvement of the closed cavity over the open cavity reflects the combined effect of near-unity coupling efficiency at critical coupling and the larger interaction volume of the thicker active layer, made possible by the spatial separation of dissipation and confinement that prevents the material absorption from limiting the cavity quality factor. Because critical coupling couples the incident field into the cavity without reflection, the same architecture maximizes the intracavity driving field for field-dependent processes in low-absorption media in the weak-conversion regime.

**Optical output characteristics of open and closed cavities**

A direct consequence of critical coupling is that the pump is almost entirely absorbed by the cavity while the harmonic, generated at a non-resonant wavelength, propagates freely. The closed cavity therefore functions as a self-filtering frequency converter, simultaneously enhancing the harmonic signal and suppressing the residual pump.

Fig. 5a compares the spatial beam profiles of the reflected FW (910 nm) and the SH signal (455 nm). The open cavity exhibits a distinct Gaussian-like reflected FW beam, while the closed cavity shows strongly suppressed FW reflection consistent with critical coupling. Consequently, the ratio of SHG intensity to residual FW intensity is approximately $1.09 \times 10^2$ times larger in the closed cavity than in the open cavity. This dramatic increase corresponds to a substantially improved signal-to-background ratio, achieved through the simultaneous suppression of the pump background and the resonant enhancement of the nonlinear emission.

The power dependence (Fig. 5b) reveals quadratic scaling at low pump powers (exponents 2.00 and 1.94 for the open and closed cavities), confirming the second-order process. Near 1 mW the closed cavity deviates significantly (exponent ~0.98), indicating saturation driven by the extreme intracavity field. Above 5 mW, the SHG intensity drops abruptly due to laser-induced damage to the top Au layer, beyond which efficient field confinement collapses[42–44]. Within the same power range, the open cavity and bare film references remain well below their damage thresholds and continue to exhibit stable quadratic scaling, highlighting the qualitative difference in intracavity field intensity between the two configurations.

The polarization dependence of the SHG emission is analyzed in Fig. 5c. All measured structures display the characteristic six-lobed polar pattern dictated by the $D_{3h}$ crystal symmetry of 3R-$MoS_2$[37,45]. The identical angular response observed across the monolayer reference, unstructured films, and both cavity architectures confirm that the vertical cavity design introduces no additional polarization anisotropy. Consequently, the angular dependence remains governed solely by the intrinsic nonlinear susceptibility of the material. Collectively, these results demonstrate that the closed cavity achieves giant nonlinear enhancement via critical coupling while simultaneously delivering a high-purity output that preserves the intrinsic spatial and polarization signatures of the 3R-$MoS_2$ layer.

**Discussion**

Our results establish critically coupled zeroth-order resonance as a cavity regime governed by the simultaneous satisfaction of phase cancellation at the metal interface and impedance matching between external coupling and absorption. In the closed cavity, the 3R-$MoS_2$ thickness sets the resonance phase while the top Au thickness independently sets the external coupling rate, spatially separating dissipation from field accumulation. This separation makes critical coupling compatible with low-absorption nonlinear media, since the same condition $\gamma_e = \gamma_i$ that minimizes reflection also maximizes the intracavity stored energy, directing the benefit of critical coupling to field-driven processes rather than to dissipation.

The experimental consequences of this design are distinctive. Zeroth-order resonance alone delivers an SHG enhancement of $6.81 \times 10^4$ relative to a monolayer reference in the open cavity. Adding critical coupling in the closed cavity nearly doubles this to $1.19 \times 10^5$ while narrowing the spectral linewidth from 44 nm to 11–27 nm, with the SHG peak tracking the reflectance minimum across a thickness-tunable range from 820 to 1030 nm. The same cavity simultaneously suppresses the reflected pump, yielding a $10^2$-fold improvement in signal-to-background ratio over the open cavity, a capability not naturally realized in metasurface architectures where the pump and harmonic share the same external coupling channel. The polarization signature of 3R-$MoS_2$ is preserved across all configurations.

From a practical standpoint, the entire device is fabricated by thin-film deposition and transfer without lithographic patterning. The film thickness controls the resonance wavelength through phase engineering, while the top mirror thickness independently sets the external coupling rate through impedance matching, providing two orthogonal design handles. The resulting harmonic emission is spatially extended across the illuminated area, in contrast to metasurface platforms that rely on localized hot spots, and the vertical architecture is compatible with large-area chemical vapor deposition for wafer-scale implementation[21,46,47].

The generality of the framework extends well beyond the 3R-$MoS_2$ system demonstrated experimentally. Analytical predictions spanning low-loss dielectrics, hybrid semiconductors, III–V platforms, and lossy layered materials across the visible to telecommunication range (Supplementary Figs. S4–S7) illustrate that the $(n, k)$ plane prescription accommodates active media with disparate optical properties through the same minimal cavity architectures. The framework is determined by the complex refractive indices of the active material and the metal at the operating wavelength, together with the layer thicknesses. This compact set of inputs permits natural extension to third-harmonic generation, difference-frequency generation, and other field-dependent processes[1,48] in low-absorption media. Critically coupled zeroth-order resonance therefore offers a versatile building block for ultrathin nonlinear and quantum photonic devices.

## Methods

### Fabrication

The open cavity sample was fabricated as follows. A substrate consisting of Si/ $SiO_2$ (280 nm) was prepared by depositing Ti (7 nm)/ Au (100 nm) using thermal evaporation. Subsequently, $Al_2O_3$ (7 nm) was deposited via atomic layer deposition (ALD). The $Al_2O_3$ layer served as a protective layer to suppress thermal damage to the Au and 3R-$MoS_2$ (HQ Graphene) during laser irradiation. Afterwards, 3R-$MoS_2$, mechanically exfoliated from bulk crystals, was transferred onto the substrate in varying thicknesses using a polydimethylsiloxane (PDMS) stamping method. To fabricate the closed cavity sample, an additional ALD process was performed on the transferred 3R-$MoS_2$ to deposit another $Al_2O_3$ (7 nm) layer. Subsequently, thermal evaporation was conducted once more to deposit Au (28 nm), forming a top metal layer and creating a vertical cavity structure.

### Atomic Force Microscope

AFM was carried out at Korea Advanced Nano Fab Center (KANC), using a XE-150 (Park Systems). The scans were performed in the non-contact mode with resonance frequency 320 kHz and scan rate 0.5 Hz (Fig. S10).

### SHG Measurement

SHG intensity and polarization dependence measurements were performed using the same optical setup (Fig. S14). A tunable femtosecond laser (Coherent, Chameleon-Ultra-II) operating in the 800-1050 nm range was used as the fundamental wavelength source, and a bandpass filter was employed to maintain the FW linewidth at 10 nm. The FW was focused onto the sample using a 100× objective lens with NA 0.70 (Mitutoyo, M Plan Apo NIR HR 100×). To ensure consistency of the measurements, the pumping power was fixed at 1 mW. A short-pass dichroic mirror (Thorlabs, DMSP 650R) and a 633 nm short-pass filter (Semrock, BSP01-633R-25) were used to separate the FW and the SH wavelengths. The FW polarization was adjusted using a Glan polarizer (Thorlabs, GL10B) and a half-wave plate (Thorlabs, SAHWP05M-700). For polarization dependence measurements, an additional Glan polarizer (Thorlabs, GTM10M-A) and half-wave plate (Thorlabs, AHWP05M-600) were placed in front of the spectrometer. Collected signals were analyzed through a spectrometer (Princeton Instrument, SpectraPro HRS-300) and a CCD camera (Princeton Instrument, PIXIS 400). Neutral density filter (Thorlabs, NDK01) was applied to prevent saturation of the detector. For LL curve measurements, the same optical setup was used, but the pumping power was varied using a variable neutral density filter. The SHG power required for calculating the conversion efficiency was directly measured using a powermeter. The transmission losses of each optical element were measured using a powermeter and used to correct the SHG signal accordingly (Table S2).

### Reflectance Measurement

Reflectance measurements were performed using a supercontinuum laser (NKT Photonics, Super-K laser) as a light source (Fig. S15). Reflectance data in the wavelength ranges of 600-1050 nm were extracted and analyzed. A beam splitter (Thorlabs, BS014) was used to separate the incident and reflected beams. The reflected light was analyzed by a spectrometer (Princeton Instrument, SpectraPro HRS-300) and a CCD camera (Princeton Instrument, PIXIS 400) to get reflectance spectra. An Au substrate was used as a reference. The sample reflectance was calculated as $R=I_{sample}/I_{ref}$.

**Analytical Modeling of Reflectance**

The Transfer Matrix Method (TMM) was employed to compute the reflectance for both the open and closed cavity structures. The models used for TMM calculations were composed as follows: Air/ 3R-$MoS_2$/ $Al_2O_3$ 7 nm/ Au 100 nm/ $SiO_2$ 280 nm for the open cavity, and Air/ Au 28 nm/ $Al_2O_3$ 7 nm/ 3R-$MoS_2$/ $Al_2O_3$ 7 nm/ Au 100 nm/ $SiO_2$ 280 nm for the closed cavity. The round-trip phase shift was also calculated using TMM. Since the 7 nm-thick $Al_2O_3$ layer is significantly thinner than the fundamental wavelength (800-1050 nm), it only introduces a minor phase delay in the reflection coefficient. Therefore, the round-trip phase shift was evaluated at the 3R-$MoS_2$ interfaces (see Supplementary Note 2).

**Data availability**

All data needed to evaluate the conclusions in the paper are present in the paper and/or the Supplementary Information.

**Acknowledgements**

This work was supported by the National Research Foundation of Korea (NRF) grants RS-2024-00335222 and RS-2025-25465325 funded by the Ministry of Science and ICT (MSIT), Republic of Korea (to S.H.G.); NRF grants RS-2023-00254920 and RS-2025-00514894 funded by MSIT, Republic of Korea (to Q.H.P.); and the Korea–US Collaborative Research Fund (KUCRF) grant RS-2024-00468463 (to Q.H.P.).

**Author contributions**

H.S.J., S.J.L., S.H.G., and Q.H.P. conceived the idea. H.S.J., S.J.L., and Y.H.J. fabricated the samples. H.S.J. and S.J.L. performed the optical measurements. H.S.J., S.J.L., and Q.H.P. conducted simulations and theoretical analysis. H.S.J., S.J.L., S.H.G. and Q.H.P. analyzed experimental and theoretical data. All authors contributed to writing and editing the manuscript.

**Competing interests**

The authors declare no competing interests.

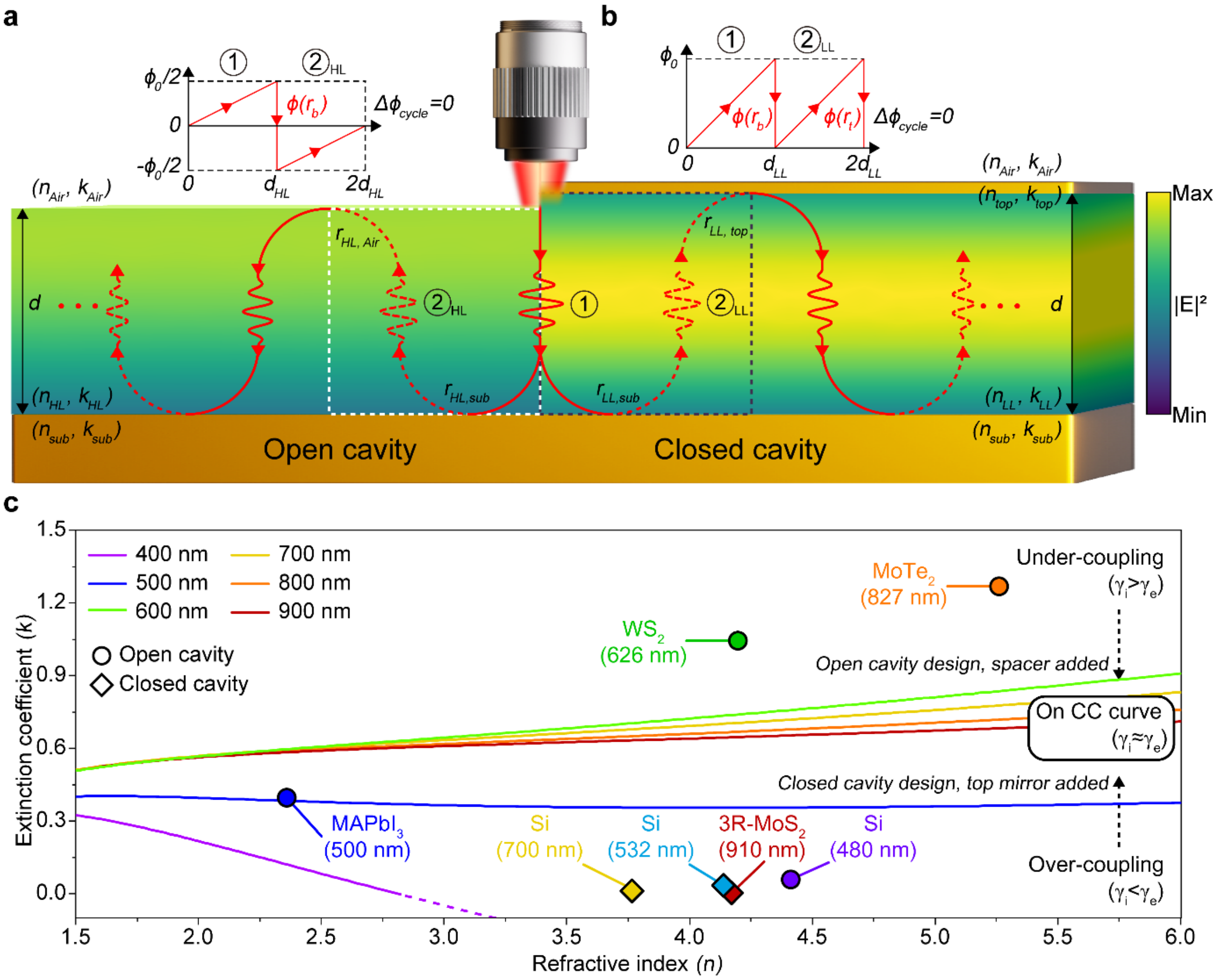

**Figure 1. Cavity operating principles of the open and closed zeroth-order cavities.** (a, b) Schematics of the open cavity (air/high-loss material/Au substrate) and the closed cavity (air/top Au mirror/low-loss material/Au substrate). In both structures, the negative reflection phase $\phi_{bot}$ at the Au interface compensates for the propagation phase accumulated in the ultrathin material layer, enabling zeroth-order resonance. In the closed cavity, the additional semi-transparent top Au mirror introduces an extra negative phase shift $\phi_{top}$ and reduces external coupling, thereby shifting the resonance to a larger thickness and enabling critical coupling of the fundamental wave. (c) Critical-coupling map in the $(n, k)$ plane showing the material combinations that satisfy the zeroth-order resonance condition and the respective coupling conditions over the wavelength range from 500 to 900 nm. Representative materials at their respective operating wavelengths are plotted to compare the under-coupled, critically coupled, and over-coupled regimes. Solid lines indicate wavelengths from 500 to 900 nm in 100 nm intervals.

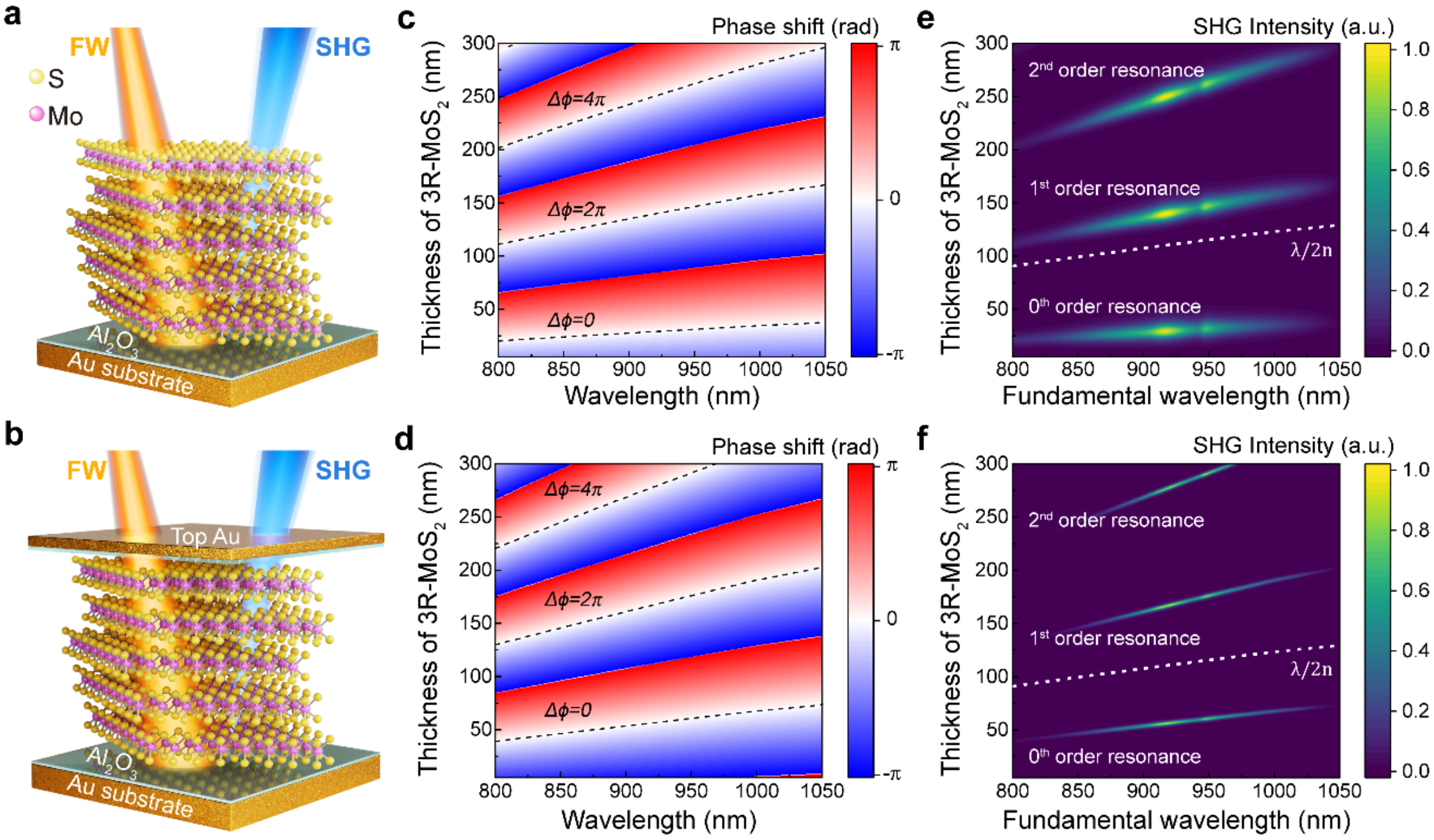

**Figure 2. Zeroth-order resonance mechanism and SHG response in the open and closed cavities.**

(a, b) Schematics of SHG in the open and closed cavities, respectively. In the open cavity, the fundamental wave (FW) is confined by zeroth-order resonance established through phase compensation at the metal interface. In the closed cavity, the semi-transparent top Au layer provides impedance matching for critical coupling while maintaining strong field confinement in the 3R-$MoS_2$ layer. (c, d) Calculated round-trip phase maps for the open and closed cavities as functions of pump wavelength and 3R-$MoS_2$ thickness. The dashed lines indicate the zeroth-order resonance branches $(\Delta\phi = 0, 2\pi, \text{and } 4\pi)$. (e, f) Corresponding numerical SHG intensity maps for the open and closed cavities. The white dashed line marks the conventional half-wavelength Fabry–Pérot thickness limit $(\lambda/2n)$.

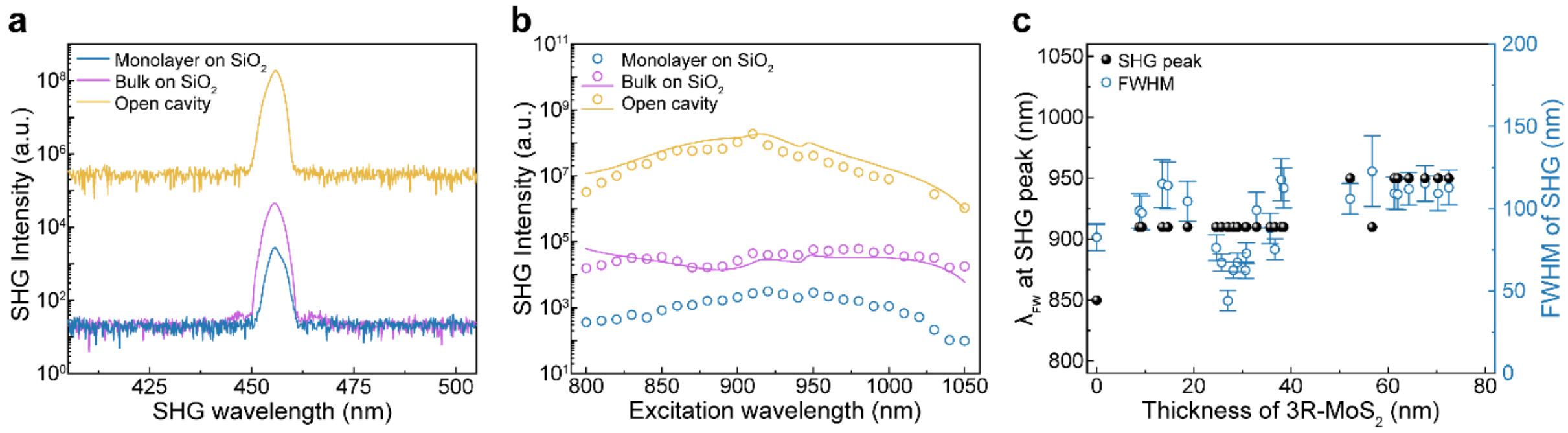

**Figure 3. Broadband SHG enhancement in the open cavity.**
(a) Comparison of SHG spectra measured from a monolayer $MoS_2$ on $SiO_2$ (blue), a 27 nm-thick 3R-$MoS_2$ on $SiO_2$ (pink), and an open cavity containing a 27 nm-thick 3R-$MoS_2$ layer (yellow). All spectra were acquired under identical excitation and collection conditions and are plotted on a common intensity scale. The open-cavity SHG spectra were corrected for the neutral density filter. (b) Excitation wavelength dependence of the SHG spectra for the open cavity containing a 27 nm-thick 3R-$MoS_2$ layer compared with the reference samples. Circles represent experimental data and solid lines indicate numerical simulations. To compare the spectral envelopes, the simulated curves are normalized to the peak value of the experimental data. (c) Thickness dependence of the nonlinear response. Black dots indicate the excitation wavelength yielding maximum SHG intensity, and blue dots represent FWHM extracted from the SHG intensity spectrum versus pump wavelength.

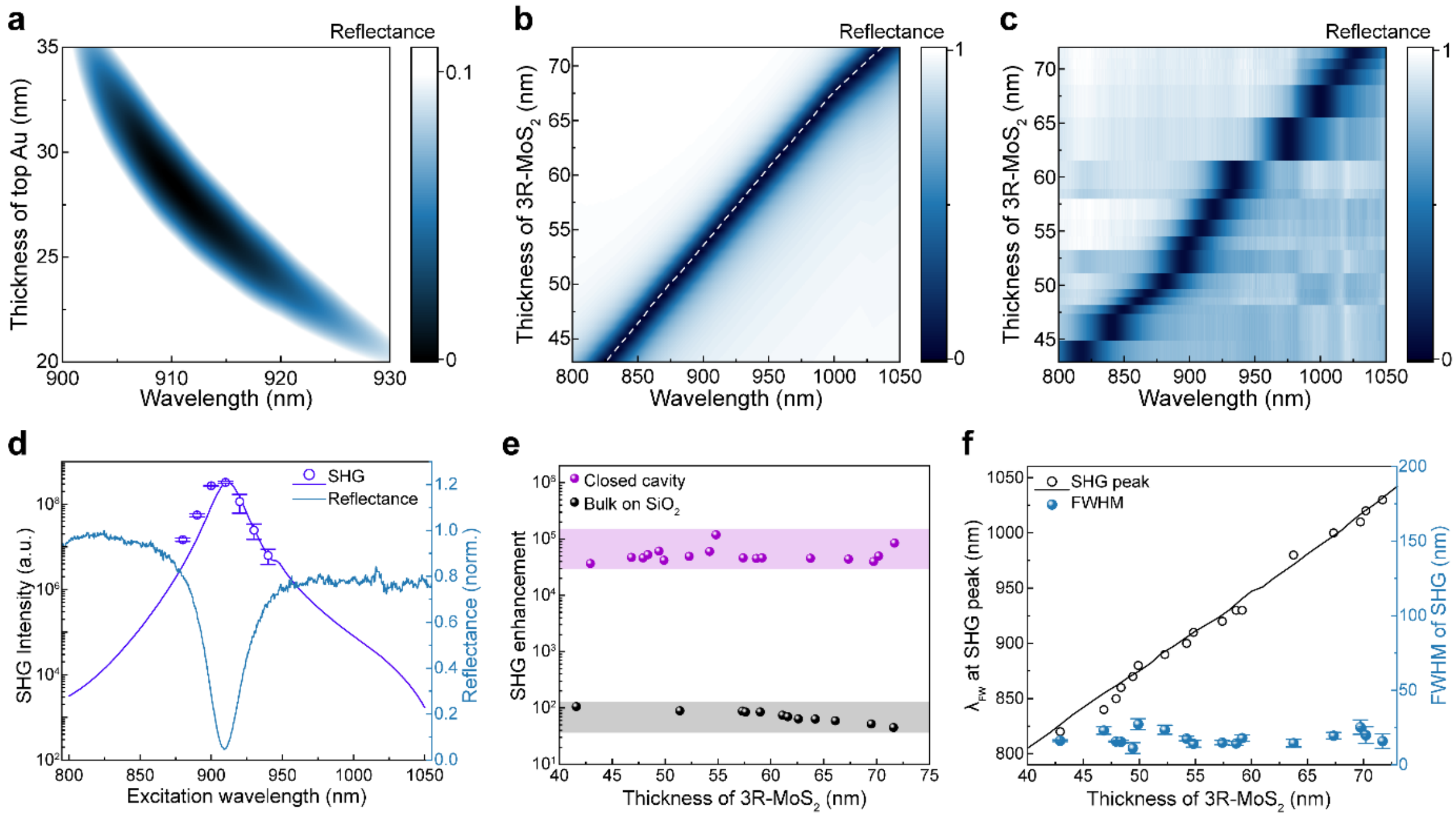

**Figure 4. Critical coupling enhanced nonlinear response in the closed cavity**

(a) Calculated reflectance map of the closed cavity as a function of wavelength and top Au thickness. The reflectance minimum indicates the critical-coupling condition achieved by impedance matching. (b) Calculated reflectance map as a function of wavelength and 3R-$MoS_2$ thickness with the top Au thickness fixed at 28 nm. The white dashed line indicates the condition under which the total round-trip phase becomes zero. (c) Corresponding experimental reflectance map of the closed cavity measured as a function of wavelength and 3R-$MoS_2$ thickness. Regions of vanishing reflectance confirm the realization of critical coupling. (d) Spectral response of a representative closed cavity sample with a 3R-$MoS_2$ thickness of 54.8 nm, showing measured reflectance (blue line), experimental SHG intensity (circles), and simulated SHG intensity (purple line). The precise alignment of the SHG peak with the reflectance dip confirms that critical coupling maximizes the nonlinear response. The global maximum occurs near 910 nm, where the cavity resonance overlaps with the peak nonlinear susceptibility of 3R-$MoS_2$. (e) SHG enhancement as a function of 3R-$MoS_2$ thickness for the closed cavity (purple dots) and unstructured 3R-$MoS_2$ film on $SiO_2$ (black dots), relative to a monolayer reference. The closed cavity achieves a maximum enhancement of $1.19\times10^5$ at 54.8 nm. Overall, the closed cavity shows an enhancement approximately three orders of magnitude larger than the unstructured films. (f) Summary of the resonance tunability. Black circles indicate the excitation wavelength of the SHG maximum and show linear tuning with thickness. Blue dots represent the spectral FWHM.

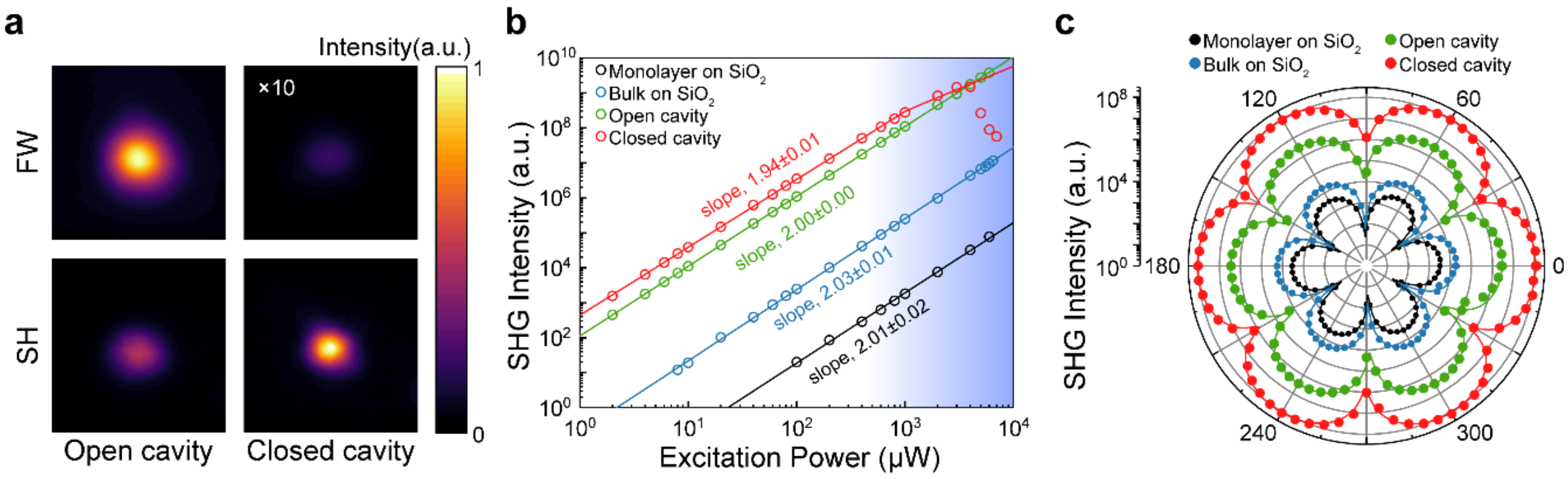

**Figure 5. Optical output characteristics: Purity, stability, and polarization.**
(a) Spatial beam profiles of the reflected FW (910 nm, top row) and the SH signal (455 nm, bottom row). The open cavity (left) exhibits strong reflection, whereas the closed cavity (right) shows suppressed reflection due to critical coupling. For visibility, the FW intensity of the closed cavity is magnified by a factor of ten; the SHG images share a common intensity scale. (b) Pump power dependence of the SHG intensity for the monolayer reference (black), unstructured film reference (blue), open cavity (green), and closed cavity (red). While the reference sample and open samples follow a quadratic scaling, the closed cavity deviates at high powers due to saturation. The blue-shaded region marks the laser-induced damage threshold where ablation occurs. (c) Polar plots of the SHG intensity versus pump polarization. All structures exhibit the characteristic six-lobed symmetry of 3R-$MoS_2$ ($D_{3h}$). Note that neutral density filters were applied to the cavity samples to prevent detector saturation.

## Supplementary Note 1. Analytical Interpretation of Critical Coupling in Ultrathin Reflective Cavities

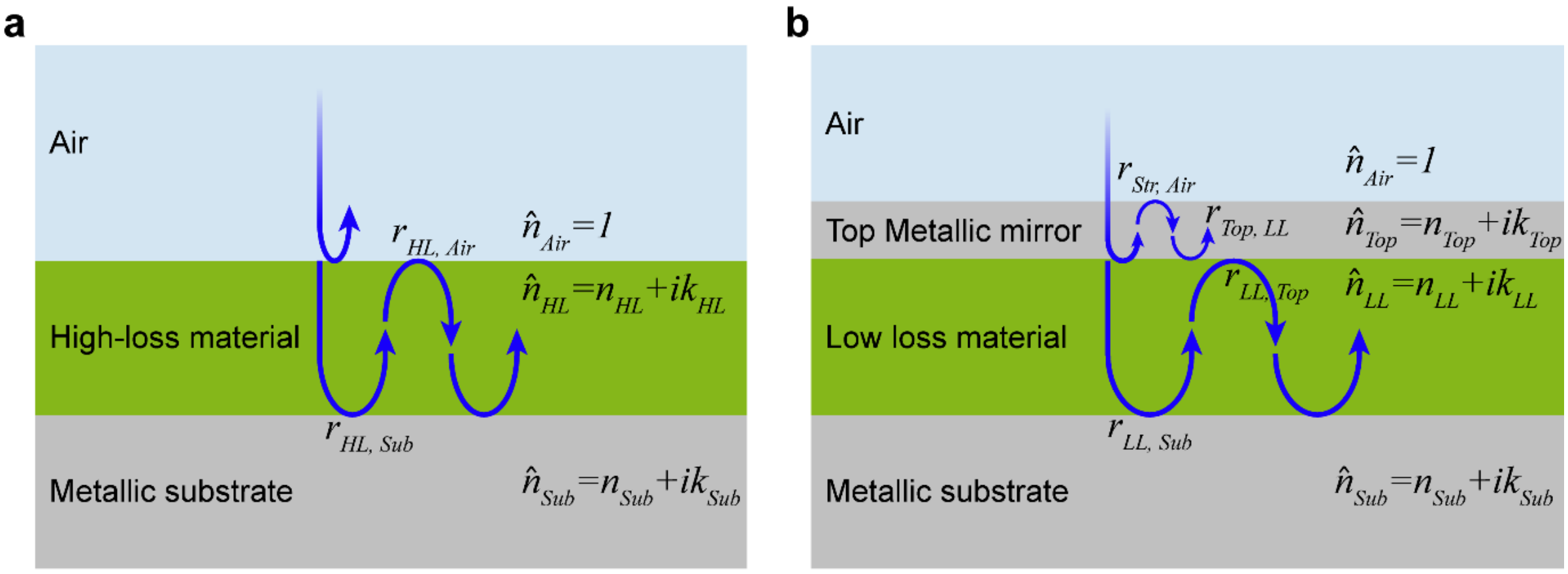

**Figure S1. Schematic illustrations of the open- and closed-cavity geometries.** (a) Open cavity (Air / high-loss material (HLM) / metallic substrate). The air–HLM interface serves as the only external coupling channel, and $r_{Air,HL}$ and $r_{HL,Sub}$ denote the reflection coefficients at the upper and lower interfaces, respectively. (b) Closed cavity (Air / top metallic mirror / low-loss material (LLM) / metallic substrate). An additional top metallic mirror of thickness $d_{Top}$ is introduced, allowing $\gamma_e$ to be tuned independently.

Within temporal coupled-mode theory, the resonant reflection of a single-port cavity is governed by the balance between the internal loss rate and the external coupling rate. For a single-port resonator with internal loss rate $\gamma_i$ and external coupling rate $\gamma_e$, the reflection coefficient at resonance is given by

$$r = \frac{\gamma_e - \gamma_i}{\gamma_e + \gamma_i} \quad \text{(N1.1)}$$

Thus, $r = 0$ when $\gamma_e = \gamma_i$, corresponding to critical coupling[1]. In a reflective cavity, $A = 1 - R$, and therefore $r = 0$ corresponds to $R = 0$ and $A = 1$.

We first consider the three-layer open cavity shown in Fig. S1a, consisting of Air / high-loss material (HLM) / metallic substrate. The open-cavity schematic in Fig. S1a represents the canonical three-layer critical-coupling geometry for high-loss films. In the experimentally relevant low-loss 3R-$MoS_2$ case, the same three-layer geometry lies in the over-coupled regime, as shown later in this note, and therefore serves only as a zeroth-order reference rather than a critically coupled cavity.

Let $r_{Air,HL}$ and $r_{HL,Sub}$ denote the reflection coefficients at the upper and lower interfaces of the HLM, respectively, and let $\phi$ denote the propagation phase delay inside the HLM. The total reflection coefficient is then

$$r = \frac{r_{Air,HL} + r_{HL,Sub}e^{-2i\phi}}{1 + r_{Air,HL}r_{HL,Sub}e^{-2i\phi}} \quad \text{(N1.2)}$$

From the reflection-cancellation condition, namely, the vanishing of the numerator, one obtains

$$e^{-2i\phi} = -\frac{r_{Air,HL}}{r_{HL,Sub}} \equiv z \quad \text{(N1.3)}$$

Defining $\hat{n}_{HL} = n_{HL} + ik_{HL}$, with layer thickness $d$ and $k_0 = 2\pi/\lambda$, gives

$$e^{2i\phi} = e^{2ik_0 n_{HL} d} \cdot e^{-2k_0 k_{HL} d} \quad \text{(N1.4)}$$

which leads to

$$d_{phase} = \frac{arg(z)}{2k_0 n_{HL}}, d_{amp} = -\frac{\ln|z|}{2k_0 k_{HL}} \quad \text{(N1.5)}$$

Under the critical-coupling condition, $d_{phase} = d_{amp}$ is required.
The round-trip phase is given by

$$\phi_{round-trip} = \phi_{prop} + \phi_{top} + \phi_{bot} \quad \text{(N1.6)}$$

and resonance occurs when

$$\phi_{round-trip} = 2m\pi (m = 0,1,2,\dots) \quad \text{(N1.7)}$$

For the zeroth-order resonance (ZOR), corresponding to $m = 0$, one has $\phi_{round-trip} = 0$, such that resonance can occur even at deep-subwavelength thicknesses($d \ll \frac{\lambda}{2n}$)[2].

The coupling coefficient is defined as

$$\kappa = \frac{\gamma_e}{\gamma_i} = \frac{Q_i}{Q_e} \quad \text{(N1.8)}$$

with

$$\frac{1}{Q_L} = \frac{1}{Q_i} + \frac{1}{Q_e}, Q_i = \frac{\omega_0}{2\gamma_i}, Q_e = \frac{\omega_0}{2\gamma_e} \quad \text{(N1.9)}$$

Accordingly, $\kappa < 1$, $\kappa > 1$, and $\kappa = 1$correspond to the under-coupled, over-coupled, and critical-coupling regimes, respectively.
Let $t_{round-trip} = 2n_{HL}d/c$, $R_{Top} =| r_{Air,HL} |^2$, and $R_{Sub} =| r_{HL,Sub} |^2$. The field evolution over one round-trip is then

$$E \to | r_{Air,HL} | \cdot | r_{HL,Sub} | \cdot e^{-2k_0 k_{HL} d} \cdot E \quad \text{(N1.10)}$$

which yields

$$\gamma_{tot} = -\frac{\ln R_{Top}}{2t_{round-trip}} - \frac{\ln R_{Sub}}{2t_{round-trip}} + \frac{4k_0 k_{HL} d}{2t_{round-trip}} \equiv \gamma_e + \gamma_i \quad \text{(N1.11)}$$

where

$$R_{Top} = \frac{(1 - n_{HL})^2 + k_{HL}^2}{(1+n_{HL})^2 + k_{HL}^2} \quad \text{(N1.12)}$$

and therefore

$$\gamma_e^{open} = -\frac{\ln R_{Top}}{2t_{round-trip}}, \gamma_i^{open} = \frac{-\ln R_{Sub} + 4k_0 k_{HL} d}{2t_{round-trip}} \quad \text{(N1.13)}$$

In the open cavity, both $\gamma_e$ and $\gamma_i$ are fixed by $(n_{HL}, k_{HL}, d)$, and therefore cannot be tuned independently. Expanding the condition $\gamma_e = \gamma_i$ yields

$$-ln\, R_{Top} = -ln\, R_{Sub} + 4k_0 k_{HL} d \quad \text{(N1.14)}$$

For low $k$ and high $n$, the left-hand side is generally much larger than the right-hand side, leading to $\kappa > 1$, that is, the over-coupled regime. Therefore, for low-loss, high-index materials such as 3R-$MoS_2$ ($k \ll 1$), critical coupling cannot be achieved in the open-cavity geometry alone, and a closed-cavity structure with an independently tunable top metallic mirror is required.

We next consider the closed cavity shown in Fig. S1b, consisting of Air / top metallic mirror / low-loss material (LLM) / metallic substrate. The top metallic mirror has refractive index $\hat{n}_{Top} = n_{Top} + ik_{Top}$ and thickness $d_{Top}$. For the LLM with $\hat{n}_{LL} = n_{LL} + ik_{LL}$ and thickness $d_{LL}$, the reflection and transmission coefficients of the top metallic mirror are given by

$$r_{Top}^{closed} = \frac{r_{LL,Top} + r_{Top,Air} e^{-2i\phi_{Top}}}{1 + r_{LL,Top} r_{Top,Air} e^{-2i\phi_{Top}}} \quad \text{(N1.15)}$$

$$t_{Top}^{closed} = \frac{t_{LL,Top} t_{Top,Air} e^{-i\phi_{Top}}}{1 + r_{LL,Top} r_{Top,Air} e^{-2i\phi_{Top}}} \quad \text{(N1.16)}$$

where $\phi_{Top} = \hat{n}_{Top} k_0 d_{Top}$. The decay rates are then

$$\gamma_e^{closed} = -\frac{\ln | r_{Top}^{closed} |^2}{2t_{round-trip}}, \gamma_i^{closed} = \frac{-\ln R_{Sub} + 4k_0 k_{LL} d_{LL}}{2t_{round-trip}} \quad \text{(N1.17)}$$

with $t_{round-trip} = 2n_{LL} d_{LL}/c$. Here, $\gamma_i^{closed}$ is determined by the LLM, whereas $\gamma_e^{closed}$ can be tuned independently via $d_{Top}$. The critical-coupling condition becomes

$$-\ln | r_{Top}^{closed}(d_{Top}) |^2 = -\ln R_{Sub} + 4k_0 k_{LL} d_{LL} \quad \text{(N1.18)}$$

Following temporal coupled-mode theory[1,3], let $s_+$ denote the input field $(| s_+ |^2 = P_{in})$ and $a$ the mode amplitude $(| a |^2\text{: } stored\ energy)$. Then

$$\frac{da}{dt} = (i\omega_0 - \gamma_e - \gamma_i)a + \sqrt{2\gamma_e}\, s_+ \quad \text{(N1.19)}$$

from which the stored energy is obtained as

$$W_{stored} = \frac{2\gamma_e}{(\gamma_e + \gamma_i)^2} P_{in} \quad \text{(N1.20)}$$

At critical coupling,

$$W_{stored}^{crit} = \frac{P_{in}}{2\gamma_i} = \frac{Q_i}{2\omega_0} P_{in} \quad \text{(N1.21)}$$

and therefore

$$\frac{W_{stored}}{W_{stored}^{crit}} = \frac{4\kappa}{(1+\kappa)^2} \equiv \eta_c \quad \text{(N1.22)}$$

where $\eta_c$ is defined as the coupling coefficient for the stored energy relative to its value at critical coupling $\kappa = 1$.

If $V_m$ denotes the mode volume, the field enhancement satisfies

$$FE^2 \propto \frac{W_{stored}}{V_m} = \frac{2\gamma_e}{(\gamma_e + \gamma_i)^2} \frac{P_{in}}{V_m} \quad \text{(N1.23)}$$

$$FE \propto \frac{\sqrt{\kappa}}{1+\kappa} \cdot \frac{1}{\sqrt{\gamma_i V_m}} \quad \text{(N1.24)}$$

$$FE_{max} \propto \frac{1}{\sqrt{\gamma_i V_m}} = \sqrt{\frac{Q_i}{\omega_0 V_m}} \quad \text{(N1.25)}$$

The above analysis can be applied directly to the experimental structures investigated in this work. We use the refractive index at $\lambda = 910$nm, where the SHG intensity of 3R-$MoS_2$ is maximal, and choose each layer thickness to satisfy the ZOR condition $\left(\phi_{round-trip} = 0\right)$.

For the open cavity (Air / 3R-$MoS_2$ (27 nm) / Au substrate), the intracavity field and field enhancement are given by

$$E_{cav}^{open}(z) = \frac{t_{Air,HL}}{1 - r_{HL,Air} r_{HL,Sub} e^{2i\phi_{HL}}} \left[e^{i\hat{n}_{HL} k_0 z} + r_{HL,Sub} e^{i\hat{n}_{HL} k_0 (2d - z)}\right] E_{in} \quad \text{(N1.26)}$$

$$FE_{open}(z) = \left|\frac{t_{Air,HL}}{1 - r_{HL,Air} r_{HL,Sub} e^{2i\phi_{HL}}}\right| \times \left|e^{i\hat{n}_{HL} k_0 z} + r_{HL,Sub} e^{i\hat{n}_{HL} k_0 (2d - z)}\right| \quad \text{(N1.27)}$$

For the closed cavity (Air / Au mirror (28 nm) / 3R-$MoS_2$ (55 nm) / Au substrate), the corresponding intracavity field and field enhancement are

$$E_{cav}^{closed}(z) = \frac{t_{Top}^{closed}}{1 - r_{Top}^{closed} r_{LL,Sub} e^{2i\phi_{LL}}} \left[e^{i\hat{n}_{LL} k_0 z} + r_{LL,Sub} e^{i\hat{n}_{LL} k_0 (2d_{LL} - z)}\right] E_{in} \quad \text{(N1.28)}$$

$$FE_{closed}(z) = \left|\frac{t_{Top}^{closed}}{1 - r_{Top}^{closed} r_{LL,Sub} e^{2i\phi_{LL}}}\right| \times \left|e^{i\hat{n}_{LL} k_0 z} + r_{LL,Sub} e^{i\hat{n}_{LL} k_0 (2d_{LL} - z)}\right| \quad \text{(N1.29)}$$

The results for the two geometries are summarized in Table S1. Here, $\langle FE^2 \rangle$denotes the thickness-averaged field enhancement across the cavity layer, evaluated as $\langle FE^2 \rangle = d^{-1}\int_0^d FE\,(z)^2\,dz$.

| | **Open cavity** | **Closed cavity** |
|---|---|---|
| Architecture | Air/ 3R-$MoS_2$ (27 nm)/ Au substrate | Air/ Top Au (28 nm) mirror / 3R-$MoS_2$ (55 nm)/ Au substrate |
| $\gamma_e\ (s^{-1})$ | $6.517 \times 10^{14}$ | $1.856 \times 10^{13}$ |
| $\gamma_i\ (s^{-1})$ | $3.246 \times 10^{13}$ | $1.644 \times 10^{13}$ |
| $\kappa$ | 20.08 (Over coupling) | 1.129 (Near critical coupling) |
| $\eta_c = \frac{4\kappa}{(1+\kappa)^2}$ | 0.181 | 0.996 |
| $W_{stored}/P_{in}\ (s)$ | $2.78 \times 10^{-15}$ | $3.03 \times 10^{-14}$ |
| $Q_L$ | 1.51 | 29.6 |
| $\langle FE^2 \rangle$ | 1.646 | 0.428 |

**Table S1**. Comparison of the key performance metrics of the open and closed cavities. The decay rates, coupling coefficient $\kappa$, $W_{stored}/P_{in}$, loaded quality factor $Q_L$, average field enhancement $\langle FE^2 \rangle$ are listed.

Direct comparison of the values in Table S1 gives the ratio of the stored energies as

$$\frac{(W_{stored}/P_{in})_{closed}}{(W_{stored}/P_{in})_{open}} = \frac{3.03 \times 10^{-14}}{2.78 \times 10^{-15}} \approx 10.9 \qquad \text{(N1.30)}$$

Likewise, the ratio of the layer-averaged field enhancement is

$$\frac{\langle FE^2 \rangle_{closed}}{\langle FE^2 \rangle_{open}} = \frac{0.428}{1.646} \approx 0.260 \qquad \text{(N1.31)}$$

Thus, the open cavity exhibits a larger local field buildup, whereas the closed cavity, being much closer to critical coupling, stores substantially more energy.

The Purcell factor is defined as[4]

$$F_P = \frac{3}{4\pi^2}\left(\frac{\lambda}{n}\right)^3 \frac{Q_L}{V_m} \qquad \text{(N1.32)}$$

and the ratio of the Purcell factors for the two structures is

$$\frac{F_P^{closed}}{F_P^{open}} = 9.62 \qquad \text{(N1.33)}$$

This indicates that the closed cavity possesses a larger $Q_L$ than the open cavity and therefore provides more favourable conditions for cavity-enhanced light–matter interactions.

## Supplementary Note 2. Analytical solution for reflectance and phase shift

In this study, we analyzed the optical responses of multilayer thin-film structures configured as open cavity and closed cavity. The optical properties of these cavities are strongly influenced by complex phenomena such as interference effects and multiple reflections occurring at the interfaces between each layer. To rigorously account for these effects, we employed the transfer matrix method (TMM) for our calculations[5]. In this note, we specifically summarize the procedure used to calculate the reflectance of the structures employed in the experiments. The specific structural information of the open cavity and closed cavity used in the calculation is shown in Figs S2a and S2b.

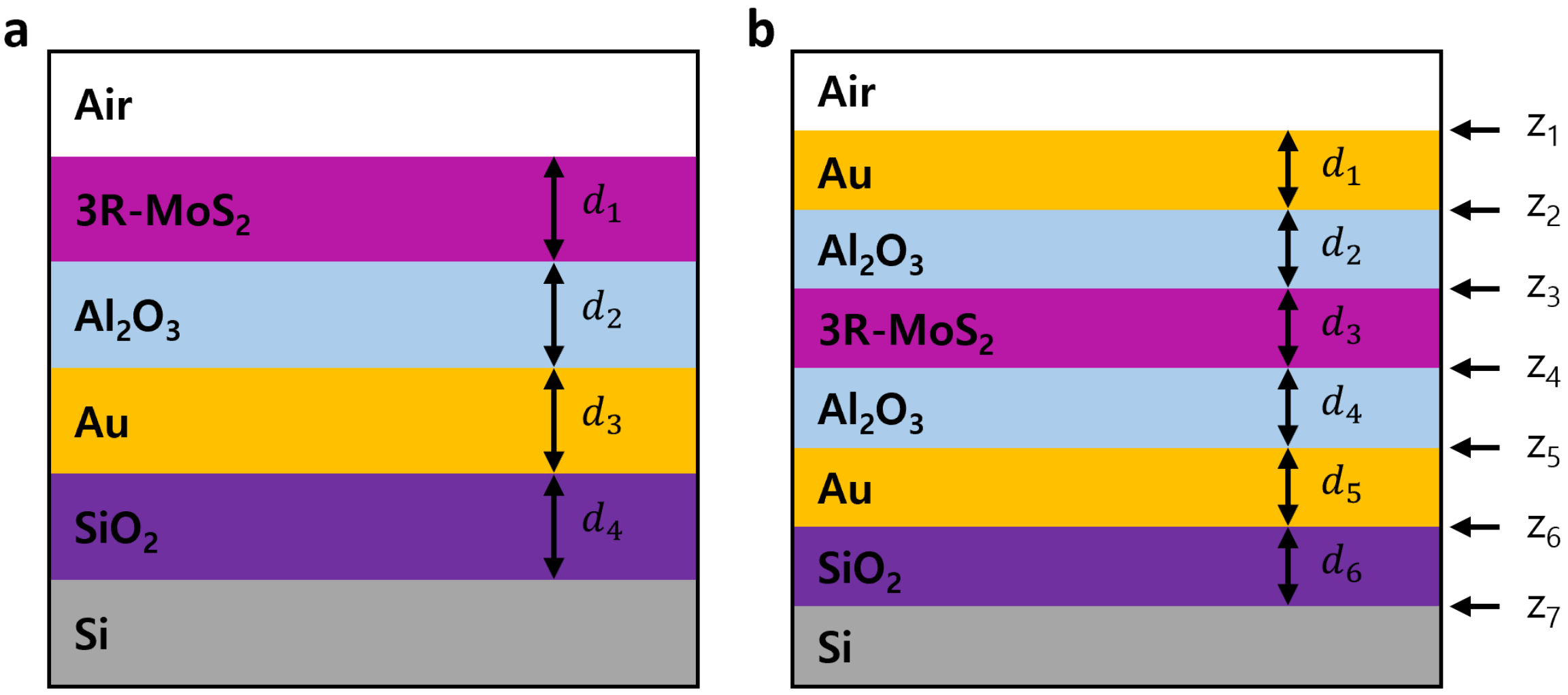

**Figure S2.** (a) Open cavity structure consisting of 6 layers(Air/ 3R-$MoS_2$ ($d_1$)/ $Al_2O_3$ ($d_2$)/ Au ($d_3$)/ $SiO_2$ ($d_4$)/ Si). (b) Closed cavity structure consisting of 8 layers(Air/ Au ($d_1$)/ $Al_2O_3$ ($d_2$)/ 3R-$MoS_2$ ($d_3$)/ $Al_2O_3$ ($d_4$)/ Au ($d_5$)/ $SiO_2$ ($d_6$)/ Si).

To analyze the cavity resonance conditions, the round-trip phase shift within each structure was also calculated using TMM. For simplicity and clear physical interpretation, all calculations were performed under the assumption of normal incidence.

The computations were carried out by inputting the complex refractive indices $n_i + jk_i$ and thicknesses of each layer into the TMM framework. From the matrix of the multilayer, the amplitude coefficients of the reflected and transmitted waves were evaluated to determine the reflectance. Under the assumption of normal incidence ($\theta_i = 0$), the dynamical matrix $D_i$ for the i-th layer is defined as follows :

$$D_i = \begin{pmatrix} 1 & 1 \\ (n_i + jk_i) & -(n_i + jk_i) \end{pmatrix} \tag{N2.1}$$

The propagation matrix $P_i$ for the i-th layer is defined as:

$$P_i = \begin{pmatrix} e^{-j\varphi_i} & 0 \\ 0 & e^{j\varphi_i} \end{pmatrix} \tag{N2.2}$$

where

$$\varphi_i = \frac{2\pi(n_i + jk_i)d_i}{\lambda} \tag{N2.3}$$

represents the phase accumulated in layer $i$ with thickness $d_i$. The total transfer matrix $M$ for the multilayer is expressed as:

$$M = \begin{pmatrix} M_{11} & M_{12} \\ M_{21} & M_{22} \end{pmatrix} = D_{Air}^{-1} \left[ \prod_i D_i P_i D_i^{-1} \right] D_{Si} \tag{N2.4}$$

The reflectance $R$ is then calculated as:

$$R = |r|^2 = \left| \frac{M_{21}}{M_{11}} \right|^2 \tag{N2.5}$$

The analytically calculated reflectance spectra for the open cavity and closed cavity structures are presented in Figs S3a and S3b, respectively.

In open cavity and closed cavity structures, the round-trip phase delay due to changes in refractive index across the top and bottom interfaces of the 3R-$MoS_2$ layer is defined as[6]:

$$\Delta\phi(\lambda, d)_{round-trip} = 2k_0 n(\lambda) d + \phi_{bot} + \phi_{top} \tag{N2.6}$$

where $k_0$ is the vacuum wavenumber, $d$ is the thickness of the 3R-$MoS_2$ layer, $n$ is its refractive index of 3R-$MoS_2$, respectively, and $\phi_{bot}$ and $\phi_{top}$ denote the reflection phase at the bottom and top interfaces of 3R-$MoS_2$. The calculated results for round-trip phase delay are represented in Figs 2c and 2d of the main text, and they are found to coincide with the positions of the reflectance minima of the same structure.

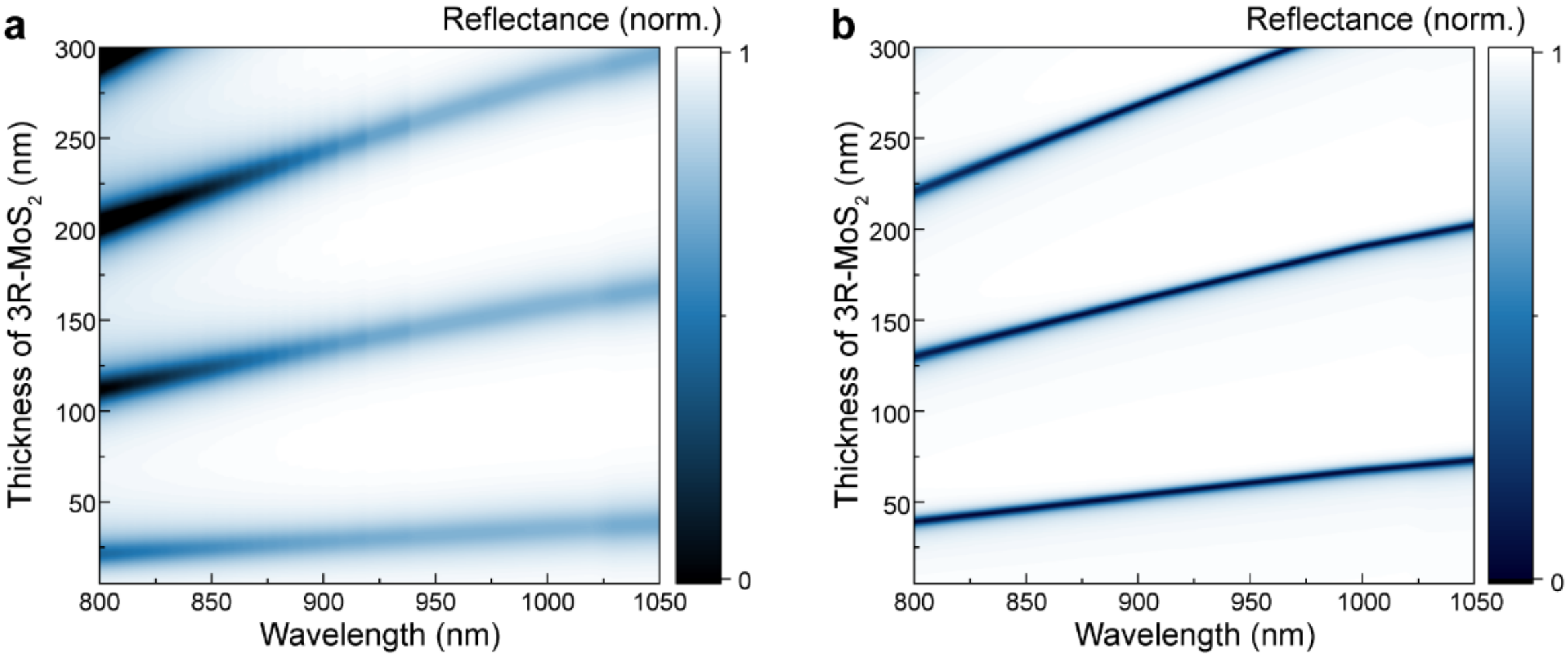

**Figure S3.** (a), (b) show the normalized analytical reflectance results for the open cavity and closed cavity, respectively. It is also observed that the regions with $\Delta\phi_{round-trip} = 0$ coincide with the reflectance minima.

## Supplementary Note 3. Numerical solution for second harmonic generation

We consider a normally incident pumping beam on a multilayer cavity embedding 3R-$MoS_2$. The multilayer structure consists of eight regions, as shown in Fig S2b, and corresponds to the closed cavity. The open cavity structure is identical to the closed cavity structure except that the topmost Au (28 nm) and $Al_2O_3$ (7 nm) layers are removed, as shown in Fig S2a. For convenience, the analysis is carried out in CGS units[7].

$$\nabla^2 \vec{E} - \frac{\varepsilon_r(\omega)}{c^2}\frac{\partial^2 \vec{E}}{\partial t^2} = \frac{4\pi}{c^2}\frac{\partial^2 \vec{P}^{NL}}{\partial t^2} \ (CGS\ units) \tag{N3.1}$$

Here, $\varepsilon_r(\omega) = 1 + \chi^{(1)}(\omega)$ denotes the linear permittivity of the nonlinear material. The nonlinear polarization vector $\vec{P}^{NL}$ is defined as follows.

$$\begin{aligned}
&\vec{E} = E_P(z)e^{-i\omega t} + E_{SH}(z)e^{-2i\omega t} + c.c. \\
&P_i^{NL} = \chi_{ijk}E_jE_k = P_i^{(1)NL}e^{-i\omega t} + P_i^{(2)NL}e^{-2i\omega t} \\
&\quad = \chi_{ijk}(\omega)(E_{SH,j}E^*_{P,k} + E^*_{P,j}E_{SH,k})e^{-i\omega t} + \chi_{ijk}(2\omega)E_{P,j}E_{P,k}e^{-2i\omega t} + c.c + [\pm 3\omega, \pm 4\omega]
\end{aligned} \tag{N3.2}$$

For a $MoS_2$ monolayer ($D_{3h}$ point group), the nonvanishing $\chi^{(2)}$ components are given by $\chi^{(2)}_{yyy} = -\chi^{(2)}_{yxx} = -\chi^{(2)}_{xxy} = -\chi^{(2)}_{xyx}$ [8]. Here, we define the $y-axis$ as the armchair direction and $x-axis$ as the zig-zag direction. For multilayered 3R-$MoS_2$(symmetry group $C_{3V}$), the nonvanishing $\chi^{(2)}$ components are $\chi^{(2)}_{zzz}, \chi^{(2)}_{xzx} = \chi^{(2)}_{yzy}, \chi^{(2)}_{xxz} = \chi^{(2)}_{yyz}, \chi^{(2)}_{zxx} = \chi^{(2)}_{zyy}$ [7,9]. Under normal incidence ($E_z^P = 0$), the dipole induced along the out-of-plane direction are negligible compared to those in the in-plane direction. Therefore, the induced nonlinear dipoles can be expressed as follows.

$$\begin{aligned}
&\chi^{(2)}_{xxy} = \chi^{(2)}_{xyx} = \alpha \\
&\chi^{(2)}_{yxx} = -\chi^{(2)}_{yyy} = \alpha \\
&\begin{cases} P_x^{(1)NL} = 2\alpha(E_{SH,x}E^*_{P,y} + E_{SH,y}E^*_{P,x}) \\ P_y^{(1)NL} = 2\alpha(E_{SH,x}E^*_{P,x} - E_{SH,y}E^*_{P,y}) \end{cases} \\
&\begin{cases} P_x^{(2)NL} = 2\alpha E_{P,x}E_{P,y} \\ P_y^{(2)NL} = \alpha(E^2_{P,x} - E^2_{P,y}) \end{cases}
\end{aligned} \tag{N3.3}$$

By substituting Eq. (N3.3) into Eq. (N3.1), we obtain the following expression.

$$\begin{aligned}
&\begin{cases} \dfrac{d^2E_{P,x}}{dz^2} + \dfrac{\varepsilon_r(\omega)\omega^2}{c^2}E_{P,x} = \dfrac{8\pi\omega^2}{c^2}\alpha(\omega)\left(E_{SH,x}E^*_{P,y} + E_{SH,y}E^*_{P,x}\right) \\ \dfrac{d^2E_{P,y}}{dz^2} + \dfrac{\varepsilon_r(\omega)\omega^2}{c^2}E_{P,y} = \dfrac{8\pi\omega^2}{c^2}\alpha(\omega)\left(E_{SH,x}E^*_{P,x} - E_{SH,y}E^*_{P,y}\right) \end{cases} \\
&\begin{cases} \dfrac{d^2E_{SH,x}}{dz^2} + \dfrac{4\varepsilon_r(2\omega)\omega^2}{c^2}E_{SH,x} = \dfrac{16\pi\omega^2}{c^2}\alpha(2\omega)\left(E_{P,x}E_{P,y} + E_{P,y}E_{P,x}\right) \\ \dfrac{d^2E_{SH,y}}{dz^2} + \dfrac{4\varepsilon_r(2\omega)\omega^2}{c^2}E_{SH,y} = \dfrac{16\pi\omega^2}{c^2}\alpha(2\omega)\left(E_{P,x}E_{P,x} - E_{P,y}E_{P,y}\right) \end{cases}
\end{aligned} \tag{N3.4}$$

This is a nonlinear coupled equation which can be solved iteratively.

$$\begin{cases} \dfrac{d^2E_{P,x}^{(i)}}{dz^2}+\dfrac{\varepsilon_r(\omega)\omega^2}{c^2}E_{P,x}^{(i)}=\dfrac{8\pi\omega^2}{c^2}\alpha(\omega)\left(E_{SH,x}^{(i-1)}E_{P,y}^{(i-1)*}+E_{SH,y}^{(i-1)}E_{P,x}^{(i-1)*}\right)\equiv S_{P,x}^{(i)};i\ge 1 \\ \dfrac{d^2E_{P,y}^{(i)}}{dz^2}+\dfrac{\varepsilon_r(\omega)\omega^2}{c^2}E_{P,y}^{(i)}=\dfrac{8\pi\omega^2}{c^2}\alpha(\omega)\left(E_{SH,x}^{(i-1)}E_{P,x}^{(i-1)*}-E_{SH,y}^{(i-1)}E_{P,y}^{(i-1)*}\right)\equiv S_{P,y}^{(i)} \end{cases}$$

$$\begin{cases} \dfrac{d^2E_{SH,x}^{(i)}}{dz^2}+\dfrac{4\varepsilon_r(2\omega)\omega^2}{c^2}E_{SH,x}^{(i)}=\dfrac{16\pi\omega^2}{c^2}\alpha(2\omega)\left(E_{P,x}^{(i)}E_{P,y}^{(i)}+E_{P,y}^{(i)}E_{P,x}^{(i)}\right)\equiv S_{SH,x}^{(i)} \\ \dfrac{d^2E_{SH,y}^{(i)}}{dz^2}+\dfrac{4\varepsilon_r(2\omega)\omega^2}{c^2}E_{SH,y}^{(i)}=\dfrac{16\pi\omega^2}{c^2}\alpha(2\omega)\left(E_{P,x}^{(i)}E_{P,x}^{(i)}-E_{P,y}^{(i)}E_{P,y}^{(i)}\right)\equiv S_{SH,y}^{(i)} \end{cases} \quad \text{(N3.5)}$$

with the initial condition: $E_{P,x,y}^{(0)}=0, \quad E_{SH,x,y}^{(0)}=0$

Explicitly, for the pumping mode with $i=1$,

$$\begin{gathered} \frac{d^2E_{P,s}^{(1)}}{dz^2}+\frac{\varepsilon_r(\omega)\omega^2}{c^2}E_{P,s}^{(1)}=0\,(s=x\,or\,y), \\ E_{P,s}^{(1)}=A_{P,s}^{(1)}e^{i\beta z}+B_{P,s}^{(1)}e^{-i\beta z}, \quad \beta=\sqrt{\varepsilon_r}\frac{\omega}{c} \end{gathered} \quad \text{(N3.6)}$$

For all other cases ($i\ge 2$),

$$\begin{gathered} \frac{d^2E_{F,s}^{(i)}}{dz^2}+\frac{\varepsilon_r(\omega_F)\omega_F^2}{c^2}E_{F,s}^{(i)}=S_{F,s}^{(i)}\,(F=P\,or\,SH) \\ E_{F,s}^{(i)}=A_{F,s}^{(i)}e^{i\beta_F z}+B_{F,s}^{(i)}e^{-i\beta_F z}+\frac{1}{2i\beta_F}\int_{z_4}^{z_3}S_{F,s}^{(i)}(z_0)e^{i\beta_F|z-z'|}dz' \end{gathered} \quad \text{(N3.7)}$$

where we used

$$\left(\frac{d^2}{dz^2}+\beta_F^2\right)e^{i\beta_F|z-z'|}=2i\beta_F\delta(z-z') \quad \text{(N3.8)}$$

To evaluate pump and second harmonic generation(SHG) modes taking nonlinear depletion into account, we solve the corresponding Maxwell equation with the following ansatz:

$$E_{F,s}^{[k]}=A_{F,s}^{[k]}e^{i\beta_F^{[k]}z}+B_{F,s}^{[k]}e^{-i\beta_F^{[k]}z}+\delta_{k4}\int_{z_4}^{z_3}\frac{S_{F,s}(z')}{i\beta_F^{[4]}}e^{i\beta_F^{[4]}|z-z'|}dz';\ z_k<z<z_{k-1},\ s=x,y,\ F=P,SH$$

where

$$\beta_F^{[k]}=\varepsilon^{[k]}(\omega_F)^{1/2}\frac{\omega_F}{c},\ (\omega_P=\omega,\ \ \omega_{SH}=2\omega) \quad \text{(N3.9)}$$

The coefficients $\mathrm{A}_{F,s}^{[k]},\mathrm{B}_{F,s}^{[k]}$ are determined by the boundary conditions at each interface.

$$\begin{cases} \mathrm{A}_{F,s}^{[1]}=R_s^F,\mathrm{A}_{F,s}^{[8]}=0 \\ \mathrm{B}_{P,s}^{[1]}=E_s^{Inc},\mathrm{B}_{SH,s}^{[1]}=0,B_{F,s}^{[8]}=T_s^F \end{cases}$$

For brevity, we introduce a shorthand notation,

$$\Phi_{F,s}^{[k]}=\begin{bmatrix}A_{F,s}^{[k]}\\ B_{F,s}^{[k]}\end{bmatrix},\ U_F^{[k]}(z)=\begin{bmatrix}e^{i\beta_F^{[k]}z_k} & e^{-i\beta_F^{[k]}z_k}\\ \beta_F^{[k]}e^{i\beta_F^{[k]}z_k} & -\beta_F^{[k]}e^{-i\beta_F^{[k]}z_k}\end{bmatrix}. \tag{N3.10}$$

First, let us consider the boundary conditions at the interfaces away from the 3R-$MoS_2$. Then, the matching boundary conditions at $z=z_k;k=1,2$ or k=5,6,7 are given by

$$U_F^{[k]}\Phi_{F,s}^{[k]}=U_F^{[k+1]}\Phi_{F,s}^{[k+1]} \tag{N3.11}$$

Then, at the interfaces away from 3R-$MoS_2$,

$$\Phi_{F,s}^{[k+1]}=M_F^{[k]}\Phi_{F,s}^{[k]}:\quad k=1,2\ \text{ or }\ \text{k=5,6,7}$$
$$\Phi_{F,s}^{[1]}=\begin{bmatrix}R_{F,s}\\ E_{Inc,s}^F\end{bmatrix},\Phi_{F,s}^{[8]}=\begin{bmatrix}0\\ T_s^F\end{bmatrix},E_{Inc,s}^P=E_s^{Inc},E_{Inc,s}^{SH}=0 \tag{N3.12}$$

The boundary matching condition at the interfaces facing 3R-$MoS_2$ ($z=z_3,z=z_4$) needs caution. Inside the 3R-$MoS_2$ region, electric fields are

$$E_{F,s}^{[4]}=A_{F,s}^{[4]}e^{i\beta_F^{[4]}z}+B_{F,s}^{[4]}e^{-i\beta_F^{[4]}z}+e^{i\beta_F^{[4]}z}V_{F,s}^B(z)+e^{-i\beta_F^{[4]}z}V_{F,s}^T(z)$$
$$where\ V_{F,s}^B(z)\equiv\int_{z_4}^{z}\frac{S_{F,s}(z')}{i\beta_F^{[4]}}e^{-i\beta_F^{[4]}z'}dz',V_{F,s}^T(z)\equiv\int_{z}^{z_3}\frac{S_{F,s}(z')}{i\beta_F^{[4]}}e^{i\beta_F^{[4]}z'}dz' \tag{N3.13}$$

Then, B.C. at $z=z_3,z=z_4$ are

$$\begin{cases}M_F^{[3]}\Phi_{F,s}^{[3]}=\Phi_{F,s}^{[4]}+V_{F,s}^B(z_3)\begin{bmatrix}1\\0\end{bmatrix}\\ \left(M_F^{[4]}\right)^{-1}\Phi_{F,s}^{[5]}=\Phi_{F,s}^{[4]}+V_{F,s}^T(z_4)\begin{bmatrix}0\\1\end{bmatrix}\end{cases} \tag{N3.14}$$

Therefore, the equations can be arranged as follows

$$\begin{cases}G_F\begin{bmatrix}R_{F,s}\\ E_{Inc,s}^F\end{bmatrix}=\Phi_{F,s}^{[4]}+V_{F,s}^B(z_3)\begin{bmatrix}1\\0\end{bmatrix}\\ H_F\begin{bmatrix}0\\ T_{F,s}\end{bmatrix}=\Phi_{F,s}^{[4]}+V_{F,s}^T(z_4)\begin{bmatrix}0\\1\end{bmatrix}\end{cases} \tag{N3.15}$$
$$G_F=M_F^{[3]}M_F^{[2]}M_F^{[1]},\quad \mathrm{H}_F=\left(M_F^{[7]}M_F^{[6]}M_F^{[5]}M_F^{[4]}\right)^{-1}$$

Eliminating $\Phi$, we find

$$G_F\begin{bmatrix}R_{F,s}\\0\end{bmatrix}-H_F\begin{bmatrix}0\\T_{F,s}\end{bmatrix}=\begin{bmatrix}g_{11}^F & -h_{12}^F\\ g_{21}^F & -h_{22}^F\end{bmatrix}\begin{bmatrix}R_{F,s}\\T_{F,s}\end{bmatrix}=\begin{bmatrix}V_{F,s}^B(z_3)\\ -V_{F,s}^T(z_4)\end{bmatrix}-G_F\begin{bmatrix}0\\ E_{Inc,s}^F\end{bmatrix} \tag{N3.16}$$

Solving for $R_{F,s},T_{F,s}$ and $\Phi_{F,s}^{[4]}$,

$$\begin{bmatrix} R_{F,s} \\ T_{F,s} \end{bmatrix} = \frac{1}{(g_{11}^F h_{22}^F - g_{21}^F h_{12}^F)} \left\{ \begin{bmatrix} h_{22}^F V_{F,s}^B(z_3) + h_{12}^F V_{F,s}^T(z_4) \\ g_{21}^F V_{F,s}^B(z_3) + g_{11}^F V_{F,s}^T(z_4) \end{bmatrix} - \begin{bmatrix} h_{22}^F g_{12}^F - h_{12}^F g_{22}^F \\ g_{21}^F g_{12}^F - g_{11}^F g_{22}^F \end{bmatrix} E_{Inc,s}^F \right\}$$

$$\Phi_{F,s}^{[4]} = \begin{bmatrix} h_{12}^F T_{F,s} \\ h_{22}^F T_{F,s} \end{bmatrix} - \begin{bmatrix} 0 \\ V_{F,s}^T(z_4) \end{bmatrix} \quad \text{(N3.17)}$$

$$\begin{bmatrix} A_{F,s}^{[4]} \\ B_{F,s}^{[4]} \end{bmatrix} = \frac{1}{(g_{11}^F h_{22}^F - g_{21}^F h_{12}^F)} \begin{bmatrix} h_{12}^F (g_{21}^F V_{F,s}^B(z_3) + g_{11}^F V_{F,s}^T(z_4)) - h_{12}^F (g_{21}^F g_{12}^F - g_{11}^F g_{22}^F) E_{Inc,s}^F \\ g_{21}^F (h_{22}^F V_{F,s}^B(z_3) + h_{12}^F V_{F,s}^T(z_4)) - h_{22}^F (g_{21}^F g_{12}^F - g_{11}^F g_{22}^F) E_{Inc,s}^F \end{bmatrix}$$

Therefore, mode profiles can be arranged as follows

$$E_{F,s}^{[k]} = A_{F,s}^{[k]} e^{i\beta_F^{[k]} z} + B_{F,s}^{[k]} e^{-i\beta_F^{[k]} z}; k \neq 4, \quad z_k < z < z_{k-1}, \ s = x, y$$

$$E_{F,s}^{[4]} = A_{F,s}^{[4]} e^{i\beta_F^{[4]} z} + B_{F,s}^{[4]} e^{-i\beta_F^{[4]} z} + e^{i\beta_F^{[4]} z} V_{F,s}^B(z) + e^{-i\beta_F^{[4]} z} V_{F,s}^T(z);$$

$$= \frac{1}{(g_{11}^F h_{22}^F - g_{21}^F h_{12}^F)} \left\{ (h_{12}^F e^{i\beta_F^{[4]} z} + h_{22}^F e^{-i\beta_F^{[4]} z}) \left( g_{21}^F V_{F,s}^B(z_3) + (\frac{\beta_F^{[1]}}{\beta_F^{[4]}}) E_{Inc,s}^F \right) + h_{12}^F (g_{11}^F e^{i\beta_F^{[4]} z} + g_{21}^F e^{-i\beta_F^{[4]} z}) V_{F,s}^T(z_4)) \right\} + e^{i\beta_F^{[4]} z} V_{F,s}^B(z) + e^{-i\beta_F^{[4]} z} V_{F,s}^T(z); \quad \text{(N3.18)}$$

By Eq (N3.5), source terms $S_{x,y}^{F(i)}(z)$ are defined by

$$\begin{cases} S_{P,x}^{(i)} = \dfrac{8\pi\omega^2}{c^2} \alpha(\omega) \left( E_{SH,x}^{(i-1)} E_{P,y}^{(i-1)*} + E_{SH,y}^{(i-1)} E_{P,x}^{(i-1)*} \right) \\ S_{P,y}^{(i)} = \dfrac{8\pi\omega^2}{c^2} \alpha(\omega) \left( E_{SH,x}^{(i-1)} E_{P,x}^{(i-1)*} - E_{SH,y}^{(i-1)} E_{P,y}^{(i-1)*} \right) \end{cases}$$

$$\begin{cases} S_{SH,x}^{(i)} = \dfrac{16\pi\omega^2}{c^2} \alpha(2\omega) \left( E_{P,x}^{[4],(i)} E_{P,y}^{[4],(i)} + E_{P,y}^{[4],(i)} E_{P,x}^{[4],(i)} \right) \\ S_{SH,y}^{(i)} = \dfrac{16\pi\omega^2}{c^2} \alpha(2\omega) \left( E_{P,x}^{[4],(i)} E_{P,x}^{[4],(i)} - E_{P,y}^{[4],(i)} E_{P,y}^{[4],(i)} \right) \end{cases} \quad \text{(N3.19)}$$

Initial conditions

$$E_{P,s}^{(-1)} = 0, \ S_{P,s}^{(0)} = 0 \ (s = x, y) \quad \text{(N3.20)}$$

Then, mode profiles at the i-th iteration are

$$E_{F,s}^{[k],(i)} = A_{F,s}^{[k],(i)} e^{i\beta_F^{[k]} z} + B_{F,s}^{[k],(i)} e^{-i\beta_F^{[k]} z} + \delta_{k4} \int_0^h \frac{S_{F,s}^{(i)}(z')}{i\beta_F^{[4]}} e^{i\beta_F^{[4]}|z-z'|} dz'; \ z_k < z < z_{k-1}, \ s = x, y \quad \text{(N3.21)}$$

Based on the above analysis, the values of $\chi^{(2)}$ as a function of wavelength were calculated using the SHG intensity of the monolayer on $SiO_2$ shown in Fig S11. The calculated results for the open cavity and closed cavity structures are presented in Figs 2e and 2f of the main text, respectively. It is observed that the calculated SHG intensity coincides with the reflectance minima and the positions where the round-trip phase delay becomes zero for the corresponding structures in Fig S13.

## Supplementary Note 4. Second harmonic generation conversion efficiency

To determine the SHG conversion efficiency, the SHG signal was measured using a powermeter. The measurements were performed on a 54.8-nm-thick 3R-$MoS_2$ closed cavity sample, designed to support a fundamental resonance at 910 nm, corresponding to a second harmonic wavelength of 455 nm. Since the SHG signal passes through several optical components before reaching the powermeter, the transmission loss of each component was measured in advance and corrected accordingly. The overall transmission efficiency was obtained at a wavelength of 450 nm (close to 455 nm) by employing a broadband plasma light source (XWS-30, ISTEQ) in combination with a bandpass filter, measuring the transmittance of each optical component individually, and then calculating the cumulative product of these losses. The transmission losses of the optical components used in the measurement are summarized in Table S2. These values were used to correct the SHG power measured by the powermeter, thereby yielding the actual SHG conversion efficiency. The pump peak power was calculated by considering the hyperbolic secant pulse profile, which introduces a factor of 0.88[10]. For the second-harmonic pulse, we assume a pulse duration $\sqrt{2}$ times longer than that of the pump, which serves as an approximation for non-resonant bulk materials[11]. The directly measured conversion efficiency $\eta = P^{2\omega}/(P^{\omega})^2$ was $1.10\times10^{-7}$ ($W^{-1}$). This result demonstrates efficient SHG from an ultrathin 3R-$MoS_2$ layer enabled by the closed cavity resonance.

| Optics | Transmittance |
|---|---|
| Shortpass dichroic mirror | 94.3 (%) |
| 633 nm ShortPass filter | 94.2 (%) |
| Half wave plate | 88.8 (%) |
| Objective lens | 53.3 (%) |
| Total transmittance | 42.0 (%) |
| SHG power | |
| Measured SHG power | 5.12 (nW) |
| Corrected SHG power | 12.2 (nW) |

**Table S2.** Transmittance losses of the optical components. All values were obtained from direct measurements.

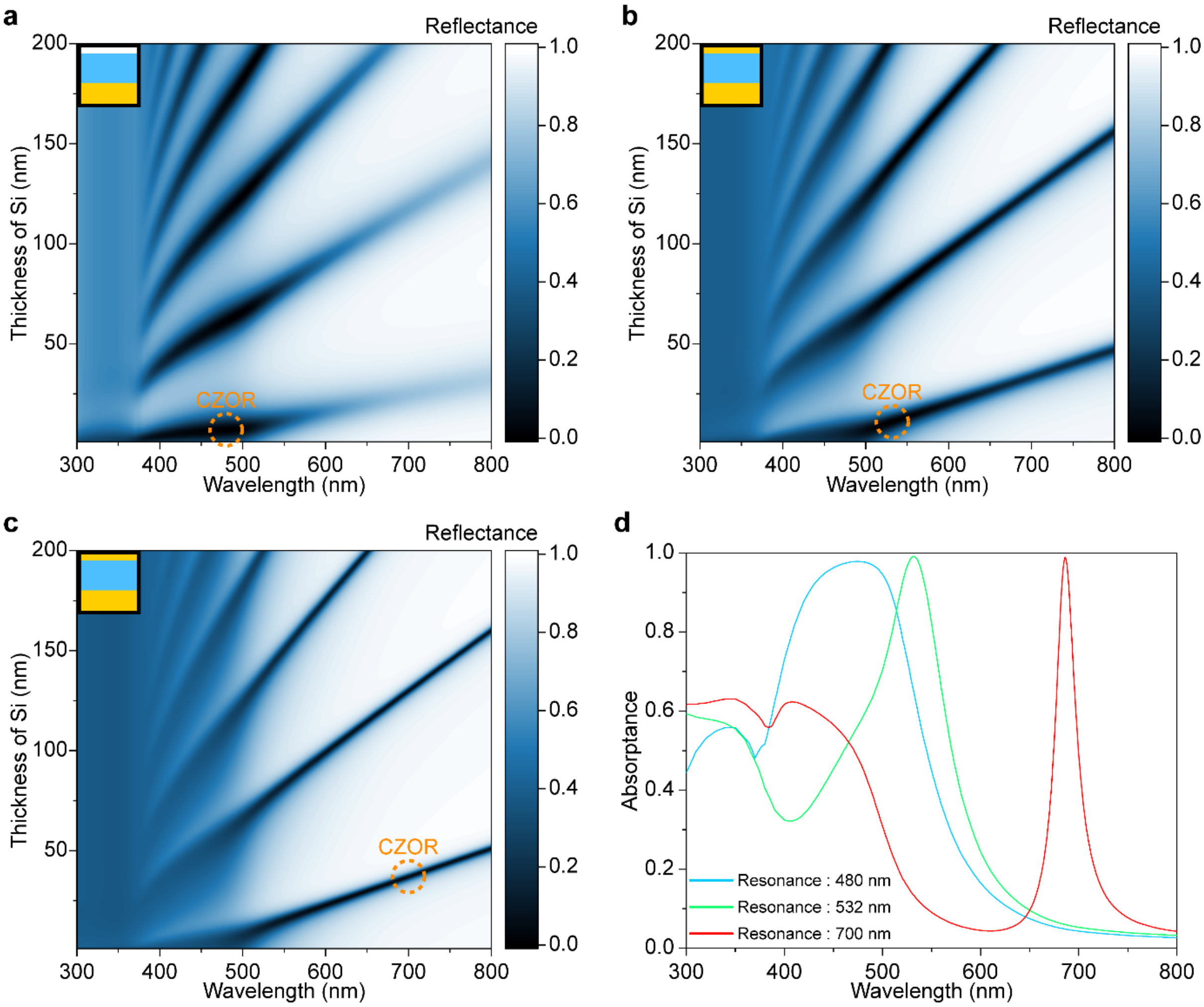

**Figure S4.** Thickness–wavelength reflectance maps of ultrathin silicon (Si)[12,13] structures for target resonance wavelengths at representative RGB wavelengths, together with the corresponding absorptance spectra under each zeroth-order resonance condition, are shown. Each reflectance map was obtained by sweeping the wavelength and Si thickness. (a) Open-cavity reflectance map for the 480 nm zeroth-order resonance condition, which is satisfied at Si = 7 nm and $Al_2O_3$ = 1 nm. (b) and (c) Closed-cavity reflectance maps for the 532 nm and 700 nm zeroth-order resonance conditions, respectively; the condition is satisfied at Si = 11 nm, $Al_2O_3$ = 10 nm, and top Au = 19 nm in (b), and at Si = 37 nm, $Al_2O_3$ = 10 nm, and top Au = 34 nm in (c). The location of the optimal zeroth-order resonance is indicated by the dashed orange circle in the reflectance map. (d) Absorptance spectra of each structure corresponding to the zeroth-order resonance conditions in (a)-(c). These results show that distinct resonances corresponding to RGB wavelengths can be realized even with a Si layer of deep-subwavelength thickness. In particular, the preservation of clear RGB resonances despite the high loss in the short-wavelength region of the visible range demonstrates that the proposed structure can simultaneously achieve extreme compactness and high spectral selectivity.

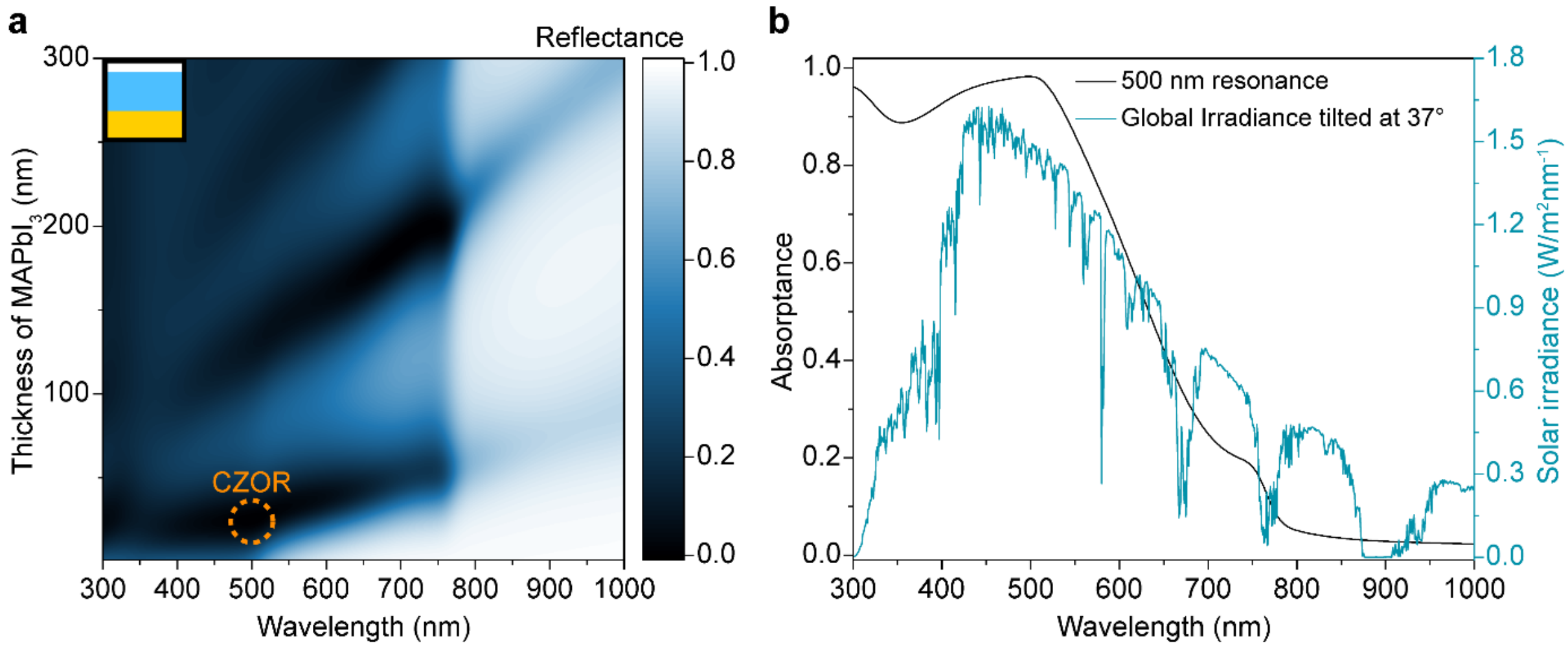

**Figure S5.** (a) Thickness–wavelength absorptance map of $MAPbI_3$[14] in the open-cavity structure under the optimized zeroth-order resonance condition at a wavelength of 500 nm. The zeroth-order resonance is satisfied at $Al_2O_3$ = 1 nm and $MAPbI_3$ = 24 nm. (b) The absorptance corresponding to the 500 nm resonance condition is presented together with the 37° tilted global solar irradiance. As a result, it is confirmed that a broad resonance can be formed over a relatively wide spectral range even in a $MAPbI_3$ layer with a thickness of only about 24 nm. This demonstrates that effective light–matter interaction is possible even in an extremely thin active layer, suggesting that such characteristics are advantageous for ultrathin light-harvesting and optoelectronic device applications.

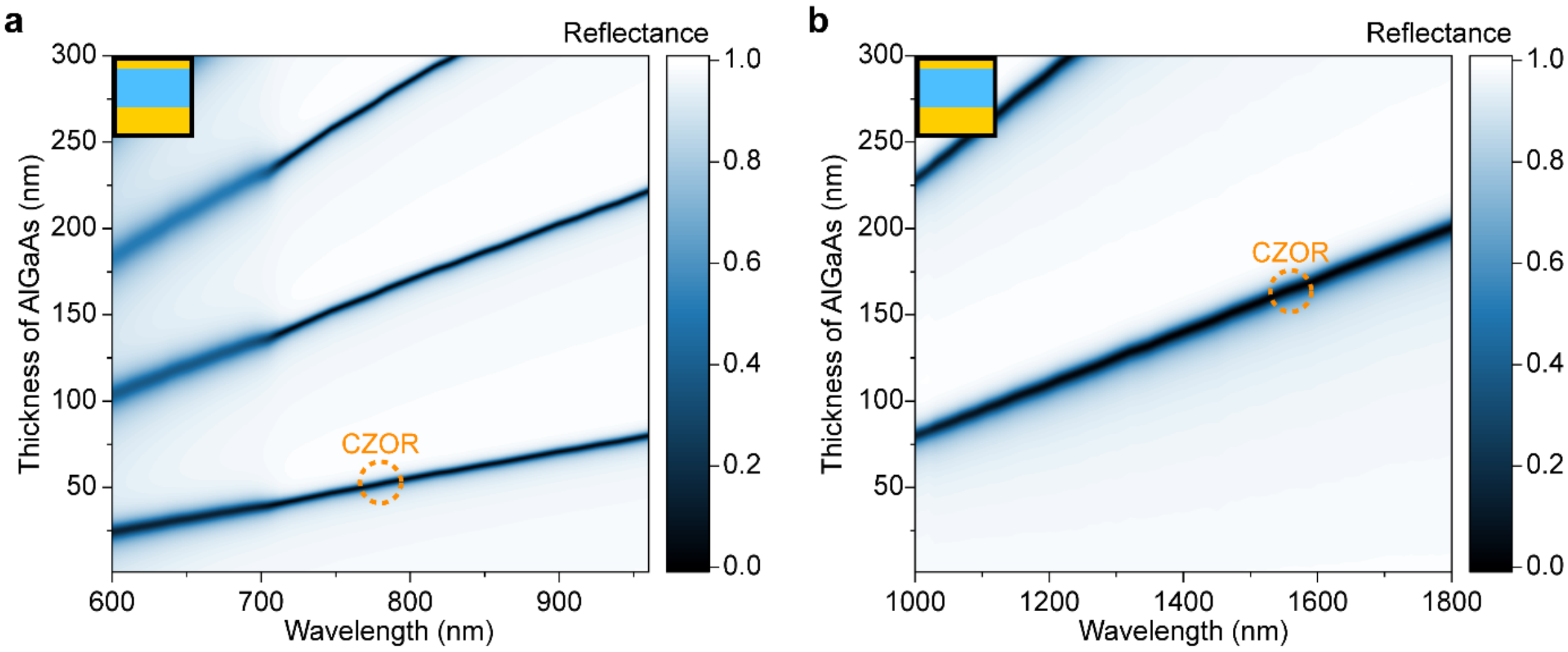

**Figure S6.** Thickness–wavelength reflectance maps of AlGaAs[15] under the optimized zeroth-order resonance conditions for 780 nm and 1560 nm are shown. (a) Closed-cavity structure, where the zeroth-order resonance is formed at 780 nm for AlGaAs = 53 nm, top Au = 31 nm, and $Al_2O_3$ = 10 nm. (b) Closed-cavity structure, where the zeroth-order resonance is formed at 1560 nm for AlGaAs = 164 nm, top Au = 17 nm, and $Al_2O_3$ = 10 nm. In each case, the optimal zeroth-order resonance condition is indicated by a dashed orange circle in the reflectance map. These results show that distinct optical responses can be realized at 780 nm and 1560 nm. In particular, the fact that 780 nm corresponds to the second harmonic of 1560 nm suggests the potential for nonlinear frequency-conversion applications. These characteristics also enable strong field confinement and enhanced light–matter interaction even in deep-subwavelength structures.

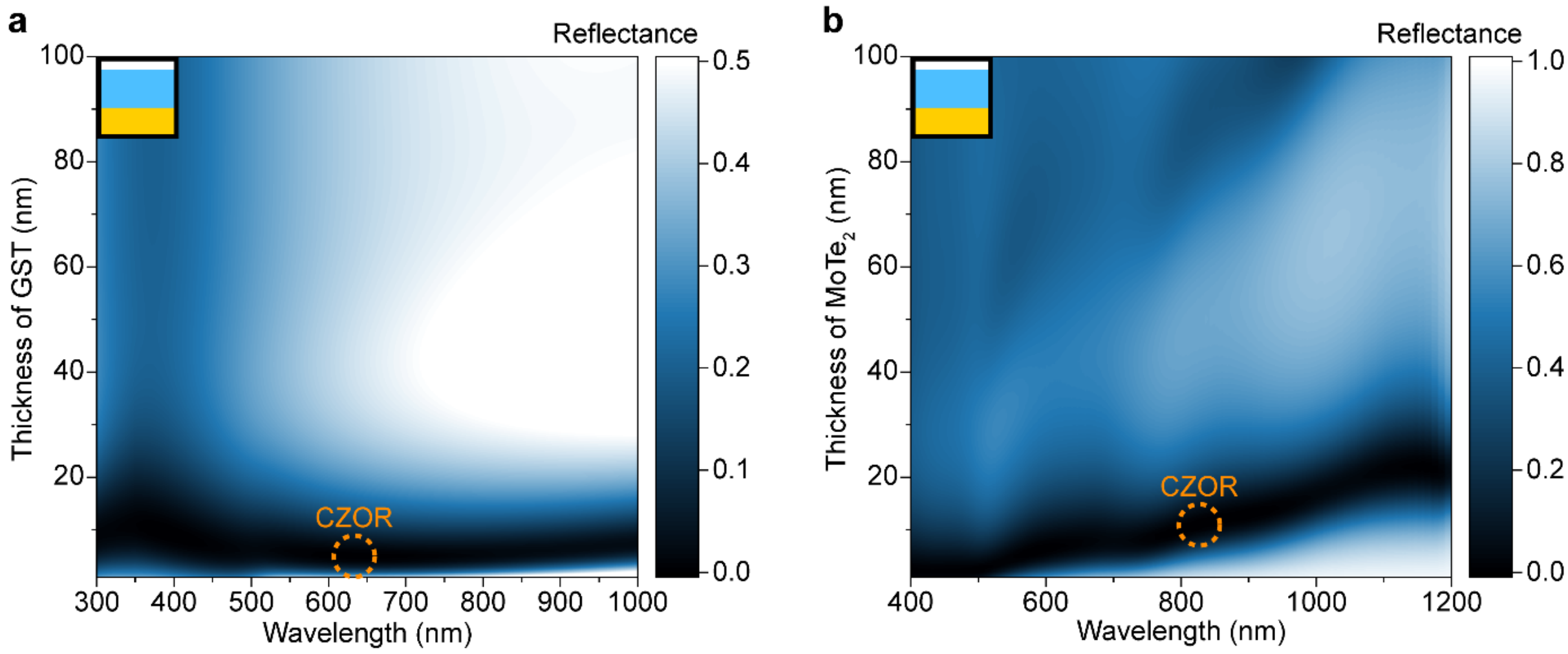

**Figure S7.** Thickness–wavelength reflectance maps under the optimized zeroth-order resonance conditions at 633 nm and 827 nm are shown for lossy GST[16] and $MoTe_2$[17] structures, respectively. (a) and (b) present the 633 nm and 827 nm zeroth-order resonances formed in the open-cavity structures of GST and $MoTe_2$, respectively; the resonance condition is satisfied at GST = 5 nm and $Al_2O_3$ = 43 nm in (a), and at $MoTe_2$ = 11 nm and $Al_2O_3$ = 21 nm in (b). Both results suggest that resonance engineering is feasible even in wavelength regions with high material loss. Therefore, these findings highlight the potential of lossy material platforms for compact photonic absorbers, spectral-selective optical components, and enhanced light–matter interaction.

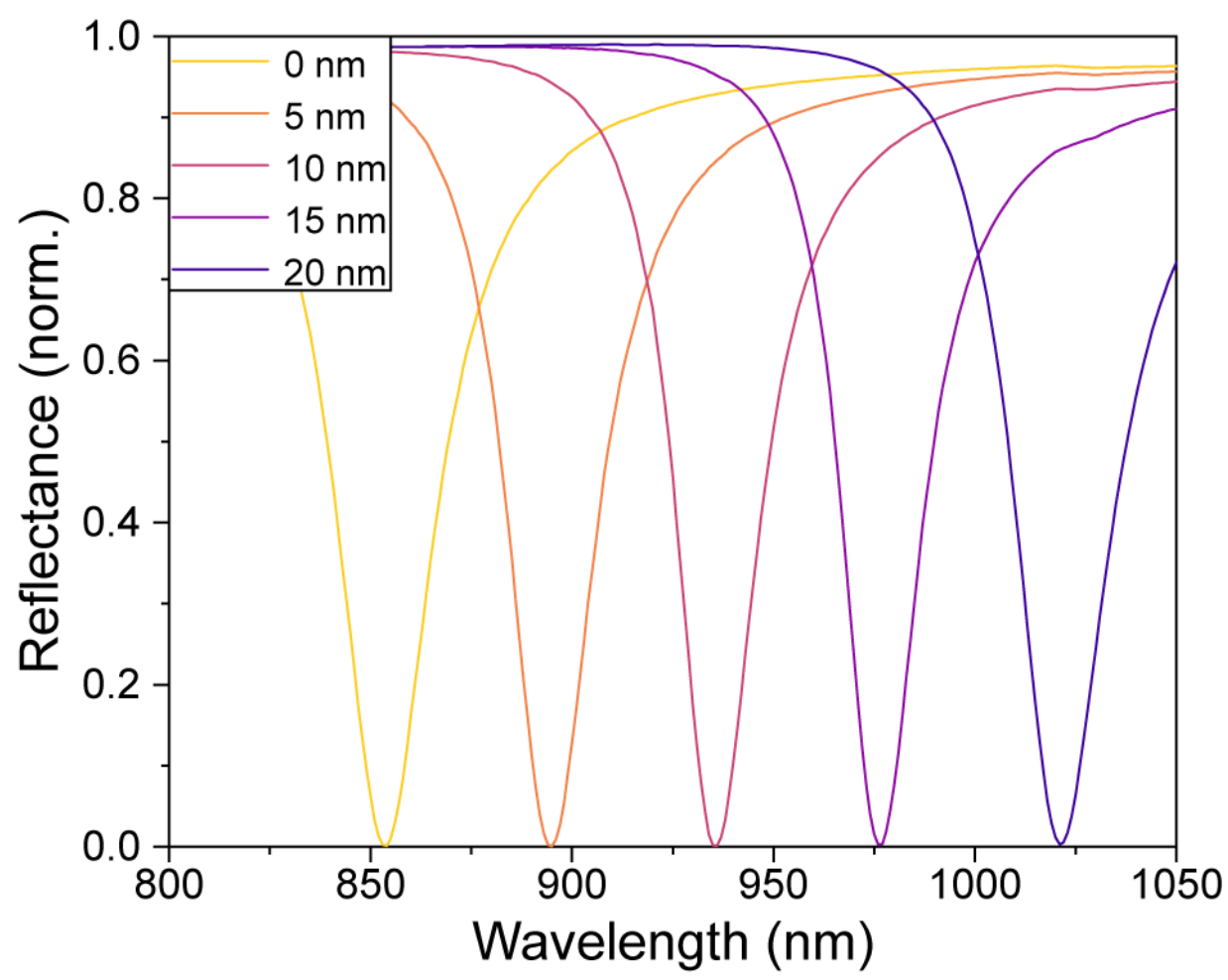

**Figure S8. ALD thickness dependent resonance tuning.** Numerical calculations of reflectance for closed cavities with the $Al_2O_3$ thickness varied from 0 to 20 nm in 5 nm steps. The thicknesses of the top Au layer and the 3R-$MoS_2$ layer are fixed at 28 nm and 55 nm, respectively.

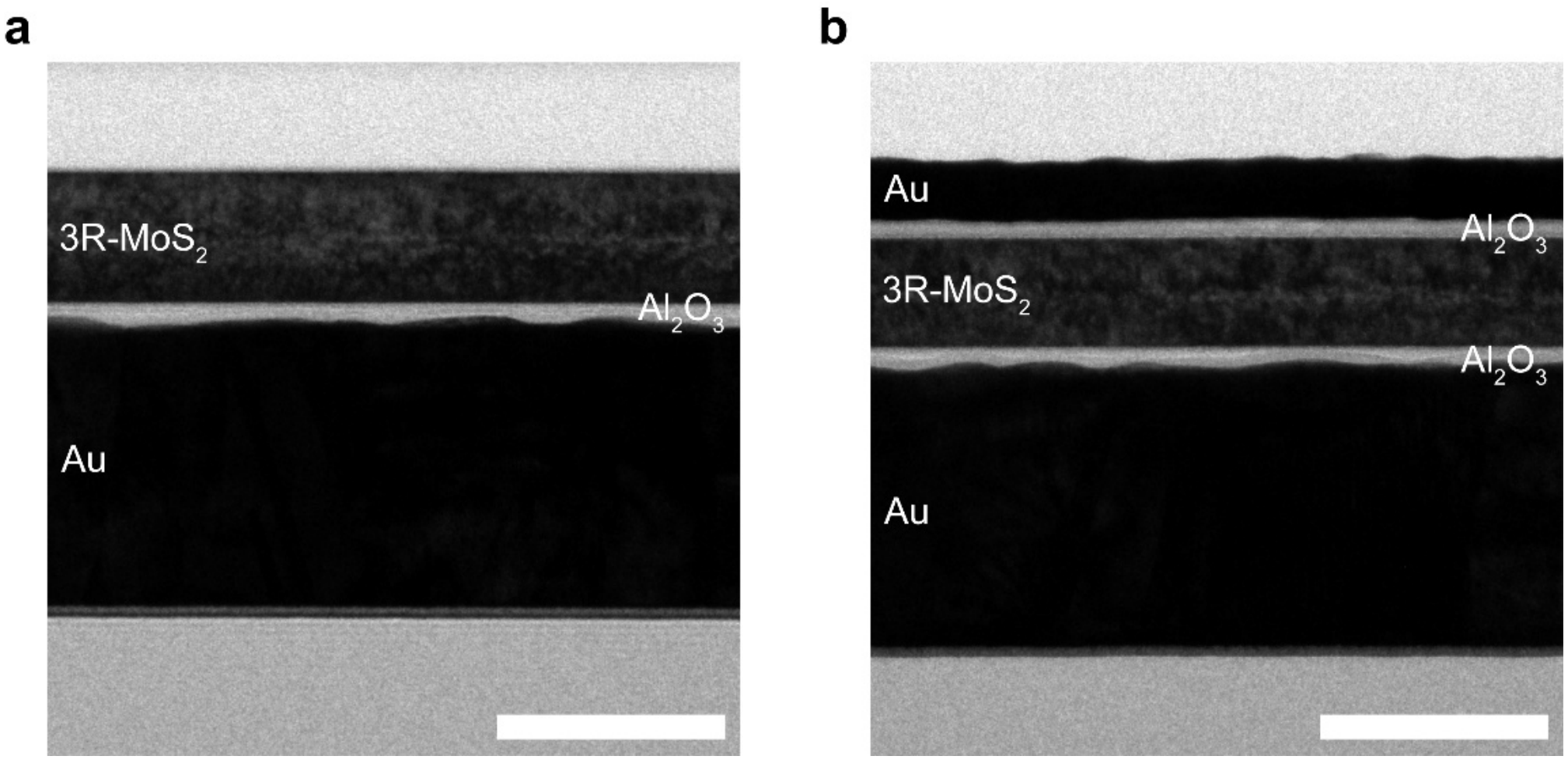

**Figure S9. Cross-sectional TEM images of the fabricated cavity structures used in the optical measurements**. (a) open cavity (3R-$MoS_2$/$Al_2O_3$/Au) and (b) closed cavity (top Au/$Al_2O_3$/3R-$MoS_2$/$Al_2O_3$/bottom Au). The vertically stacked layer sequences, including the Au layers, $Al_2O_3$ spacers, and 3R-$MoS_2$ films, are clearly resolved, confirming the successful realization of the designed structures. The $Al_2O_3$ spacer layers enable fine tuning of the optical interference phase and also protect the metal interfaces from thermal deformation and damage during optical pumping. Scale bars are 100 nm.

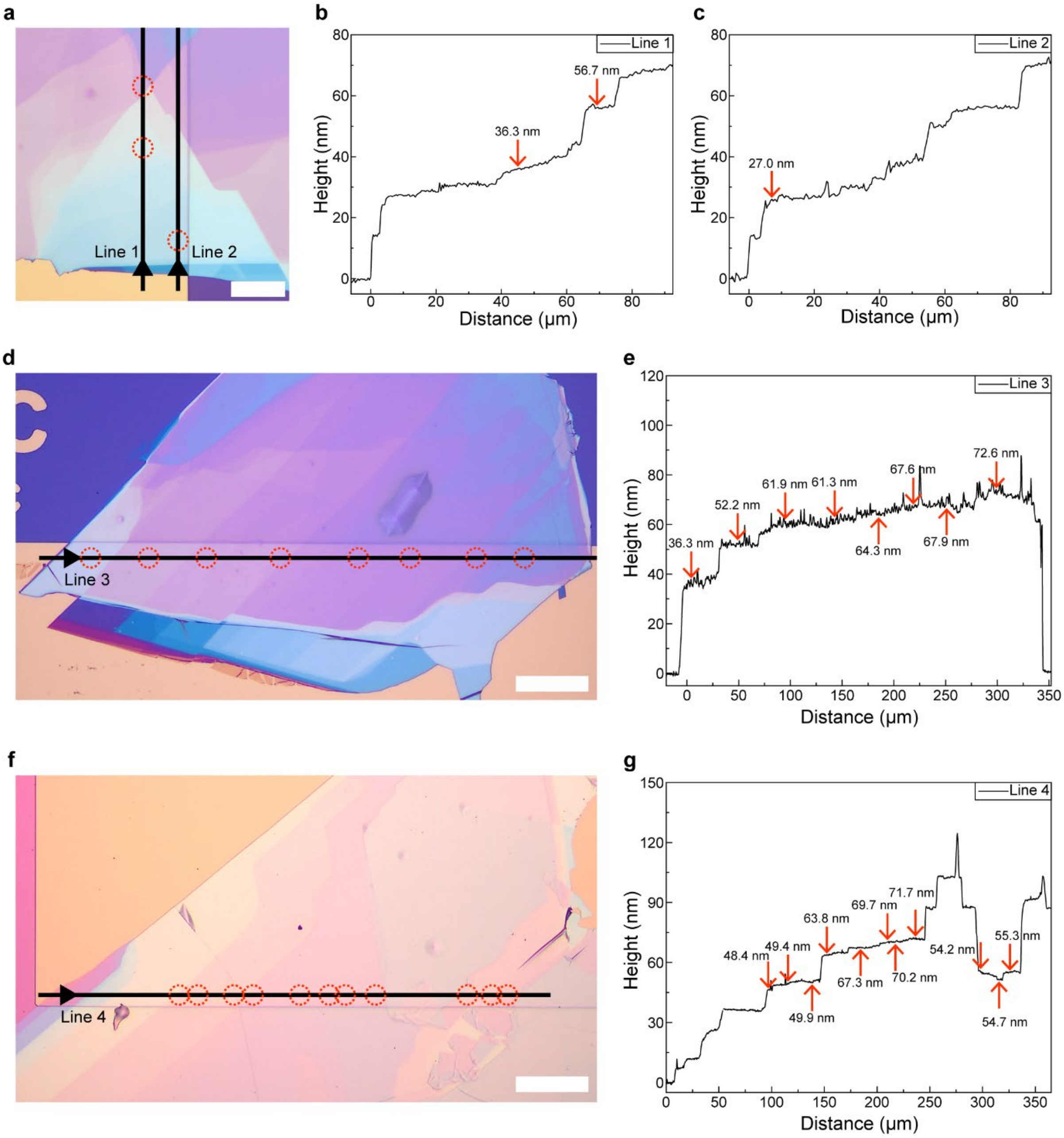

**Figure S10. Optical microscopy and AFM analysis of 3R-$MoS_2$ thickness of open and closed cavity samples.** As discussed above, the SHG intensity in both open and closed cavities is strongly influenced by the thickness of 3R-$MoS_2$. During the sample fabrication stage, the thickness of each flake was first approximately estimated using optical contrast. Subsequently, to precisely determine the thickness of the 3R-$MoS_2$ used in the experiments, the height of each sample was measured using AFM (XE-150, Park Systems). (**a**, **d**) Optical microscope images of the open cavity samples, with scale bars of 20 μm and 50 μm, respectively. The positions where the SHG intensity was measured are marked with red dashed circles, while AFM scan regions are indicated by black solid lines labeled Line 1, Line 2, and Line 3, together with the scan directions. (**b**, **c**, **e**), The corresponding AFM height profiles, respectively. (**f**) Optical microscope image of the closed cavity sample, with a scale bar of 50 μm. For the closed cavity,

the SHG measurement positions and AFM scan region are marked in the same manner as for the open cavity in (**g**).

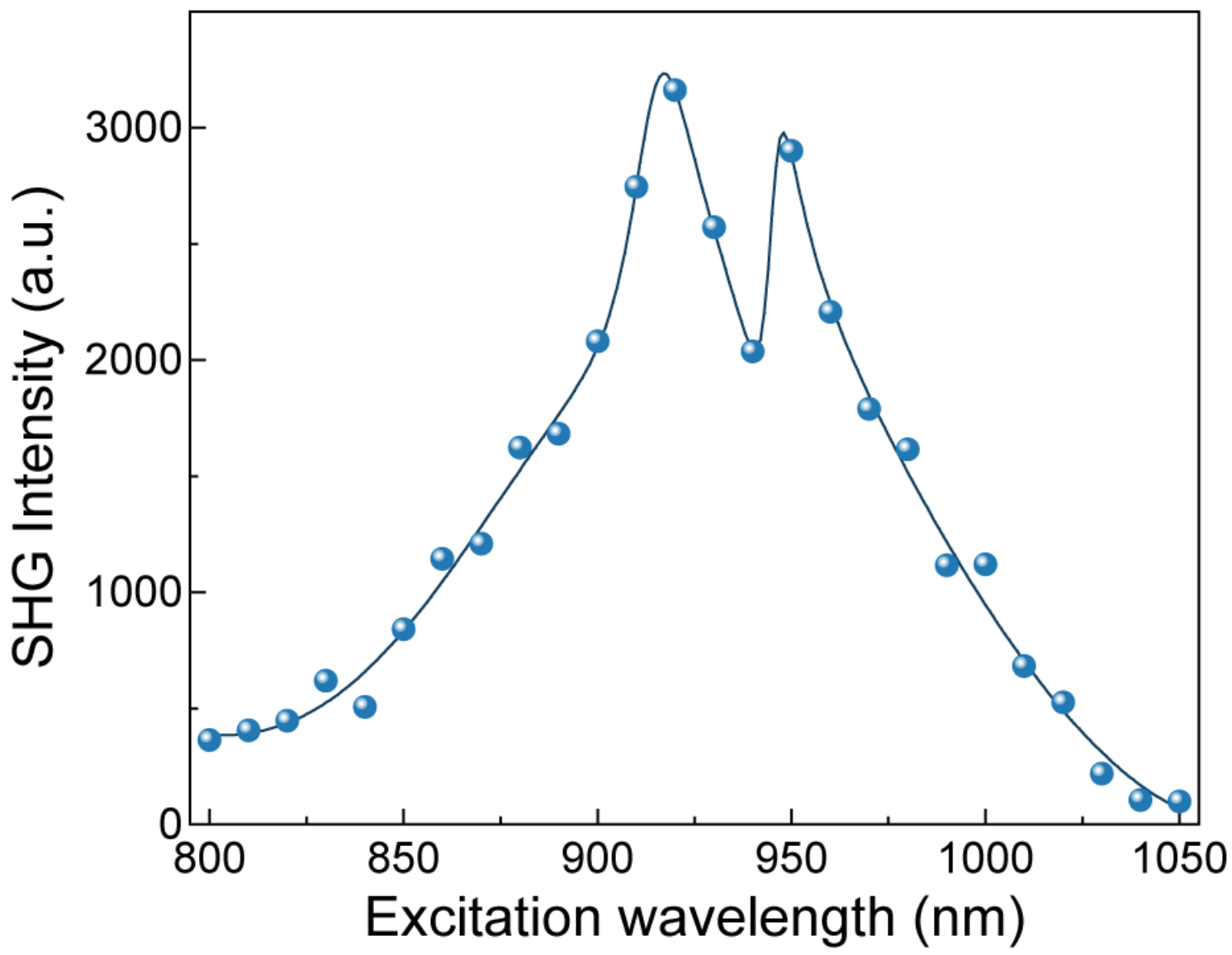

**Figure S11. Experimental results of SHG in monolayer on $SiO_2$.** The SHG measurement results (blue dot) and fitting line (blue solid line) for a $MoS_2$ monolayer on $SiO_2$ are shown. As in the open cavity and closed cavity cases, the excitation wavelength was varied from 800 to 1050 nm. The measured SHG peaks are strongly influenced by the intrinsic spectral dependence of the nonlinear susceptibility of 3R-$MoS_2$, which exhibits a dominant maximum at 910-920 nm and a secondary peak around 950 nm.

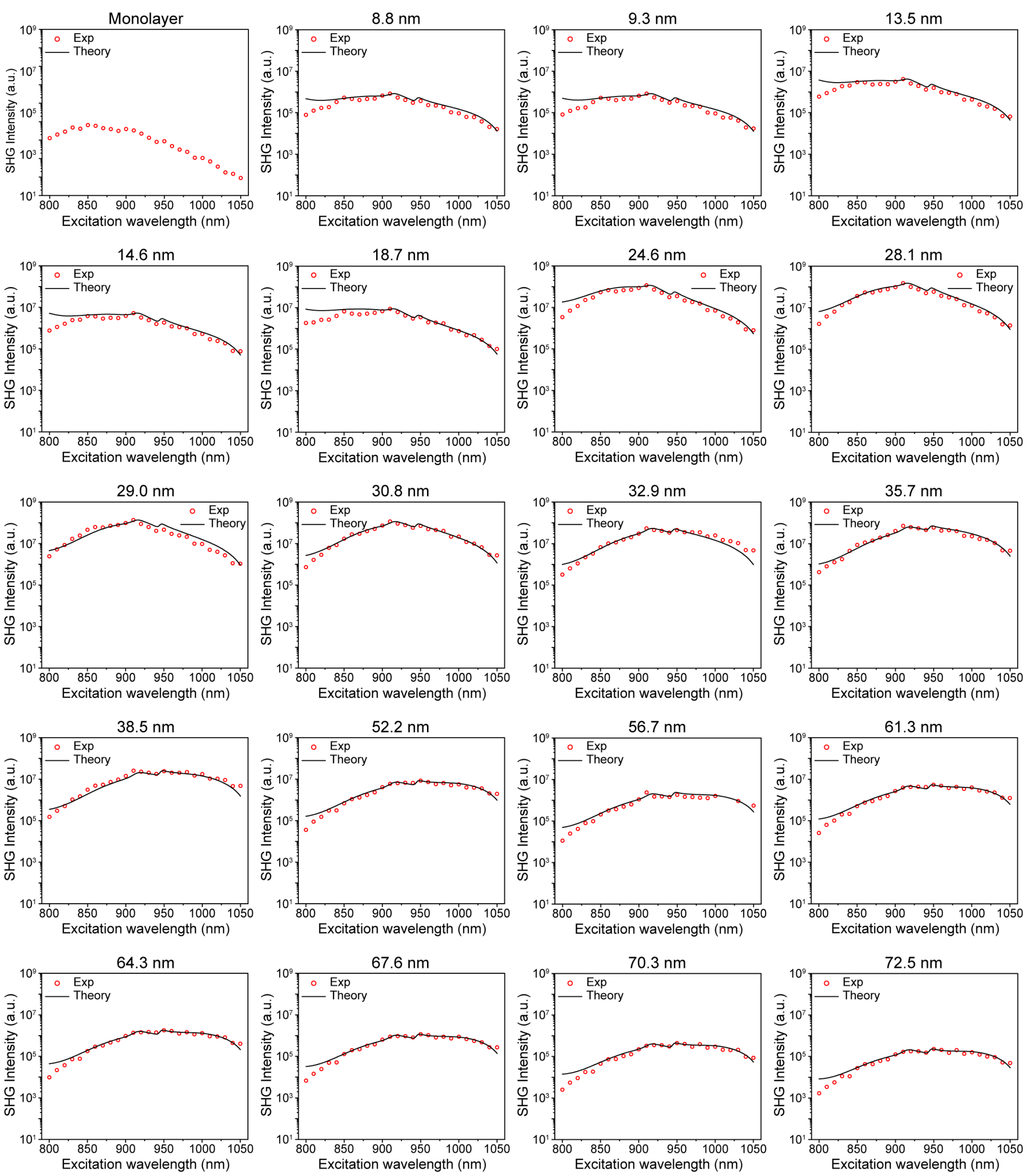

**Figure S12. Experimental and simulated results of SHG in open cavities.** Second harmonic generation (SHG) measurement results for all open cavity samples. The excitation wavelength varied from 800 to 1050 nm. The thickness of the 3R-$MoS_2$ used in each cavity is indicated at the top of each graph. Black solid lines represent numerical simulations, and red dots denote experimental data. The simulations are normalized to the peak value of the experimental results.

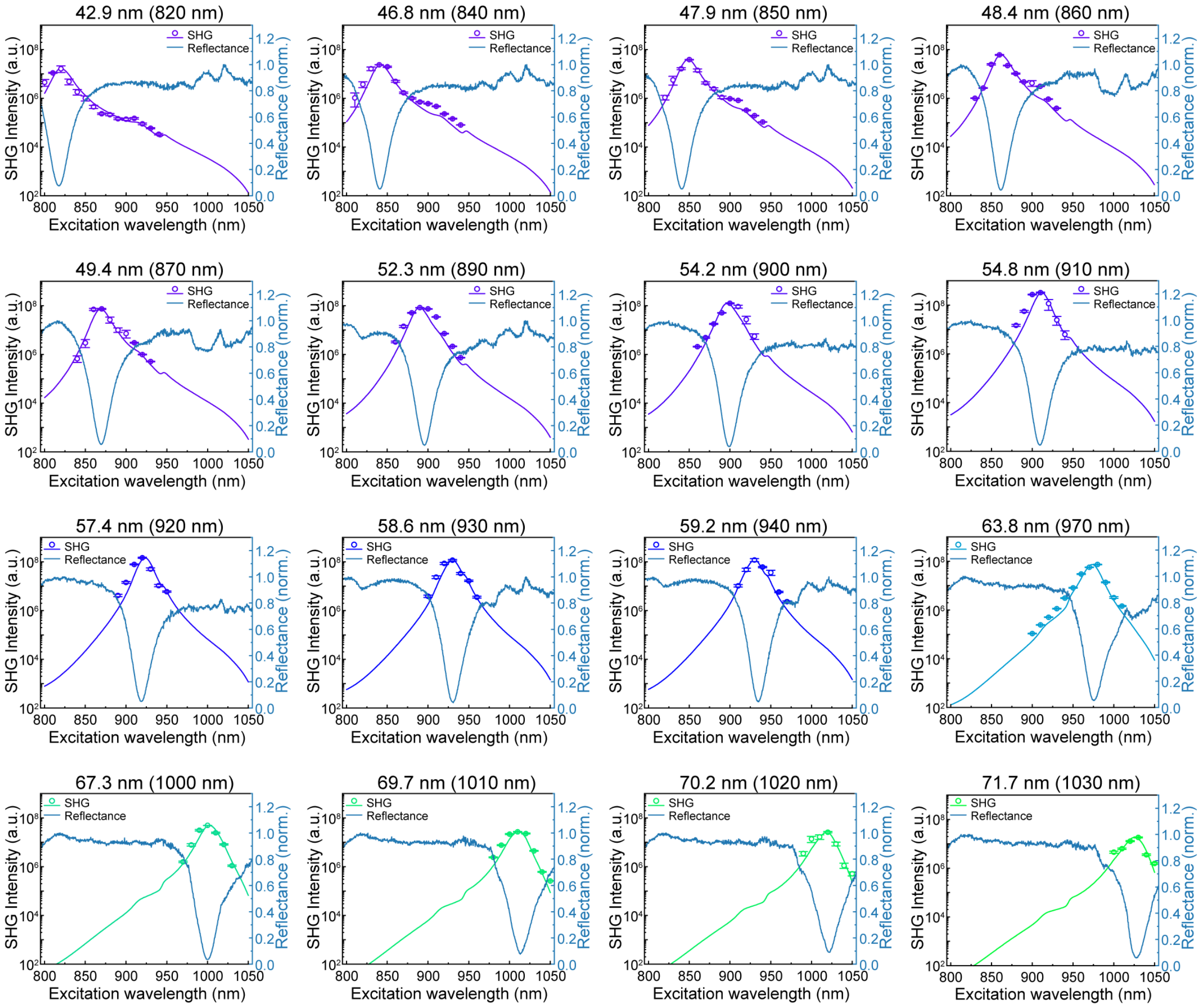

**Figure S13. Experimental and simulated results of SHG in closed cavities.** SHG and reflectance measurement results for all closed cavity samples. As in the open cavity case, the excitation wavelength varied from 800 to 1050 nm. The wavelength of the reflectance minimum coincides with the wavelength at which the SHG intensity reaches its maximum. The thickness of the 3R-$MoS_2$ used in each closed cavity sample is indicated at the top of each graph, with the corresponding resonance position given in parentheses. Solid lines (purple to green) represent SHG numerical simulations, dots (purple to green) denote SHG experimental data, and the light-blue solid line corresponds to the normalized experimental reflectance. The SHG simulations are normalized to the peak value of the experimental results.

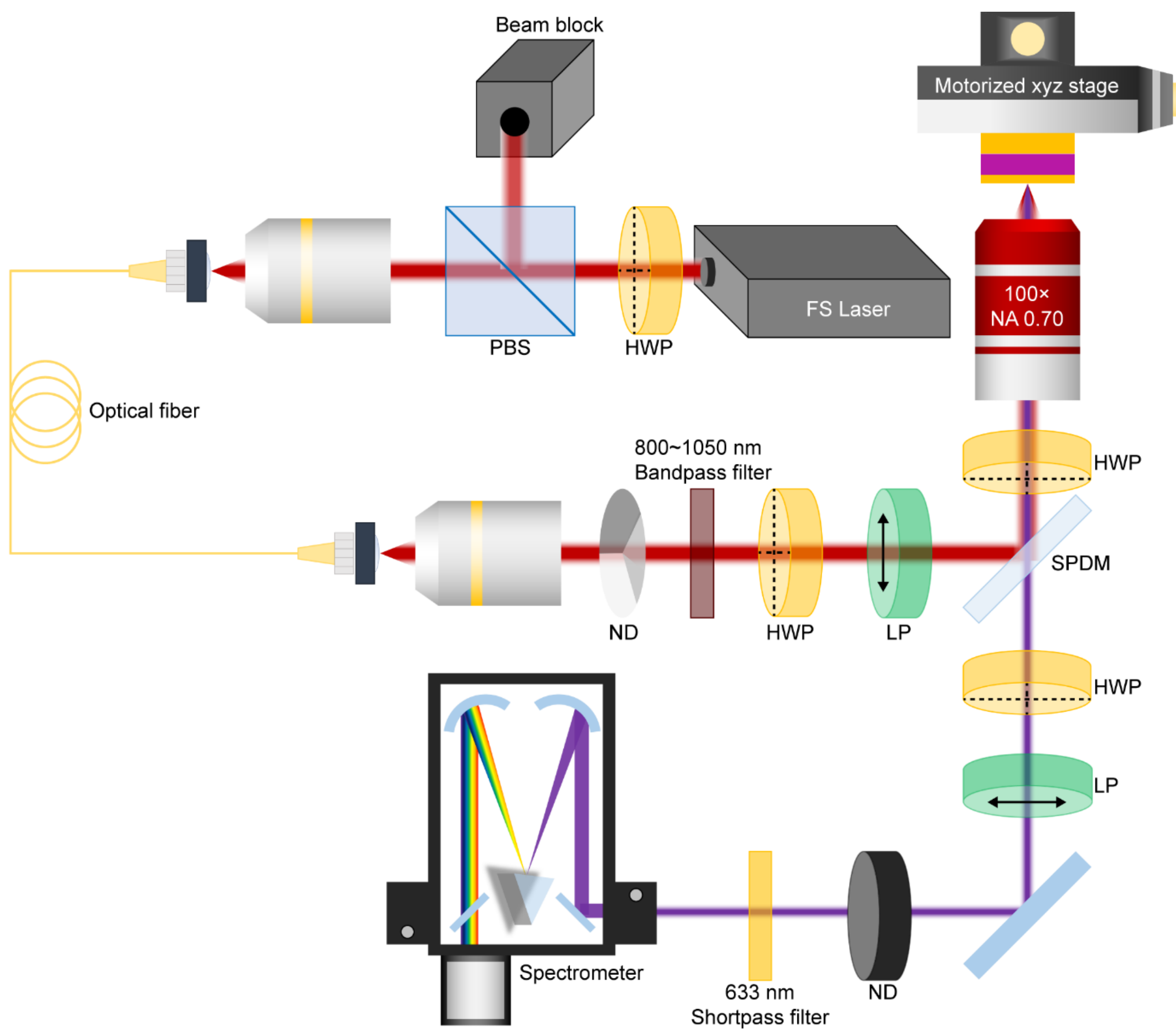

**Figure S14. Experimental setups for SHG measurements.** Experimental setup for second-harmonic generation measurement. FS laser is the femtosecond laser, HWP is the half-wave plate, PBS is the polarizing beam splitter, ND is the neutral density filter, LP is the linear polarizer and SPDM is the shortpass dichroic mirror.

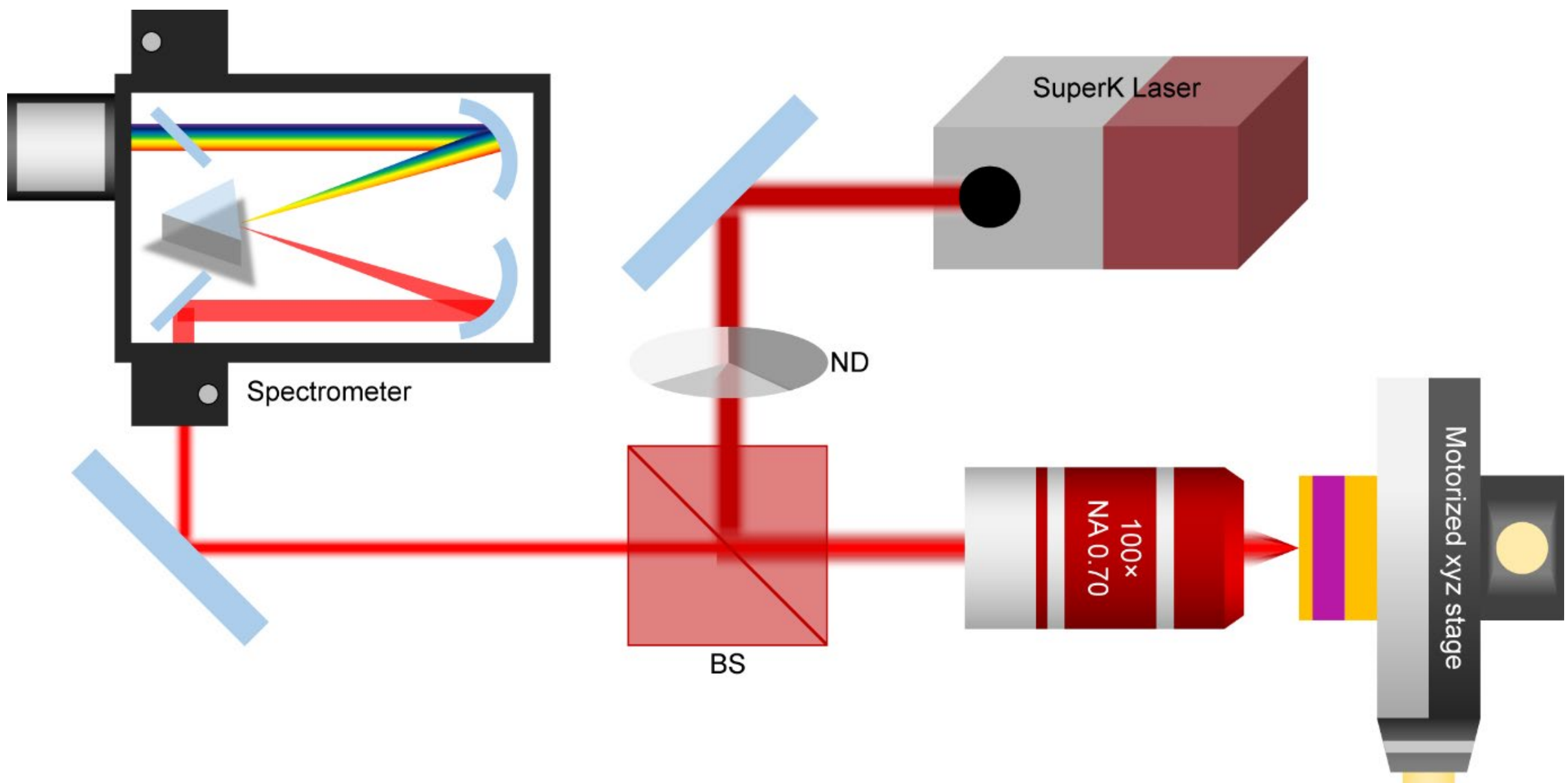

**Figure S15. Experimental setups for reflectance measurements.** Experimental setup for reflectance measurement. SuperK Laser is a broadband white-light source. ND is the neutral density filter and BS is the non-polarizing beamsplitter.

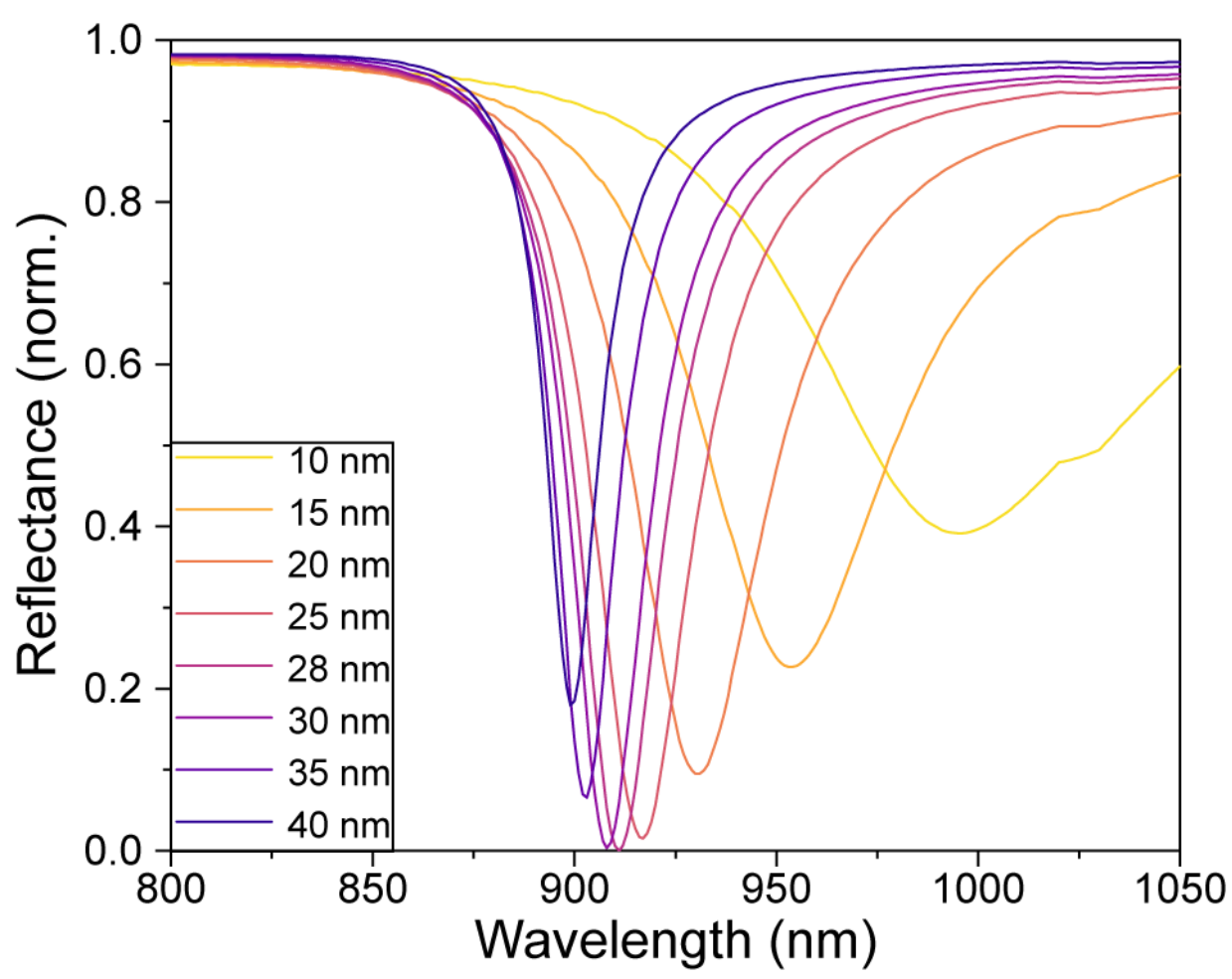

**Figure S16. Top Au thickness dependent critical coupling in closed cavities.** Numerical calculations of reflectance for closed cavities with the top Au thickness varied from 10 to 40 nm. The thicknesses of the $Al_2O_3$ layer and the 3R-$MoS_2$ layer are fixed at 7 nm and 55 nm, respectively. Critical coupling is observed when the top Au layer thickness is 28 nm.

**Supplementary References**